\documentclass[letterpaper,journal]{IEEEtran}
\usepackage{amsmath,amsfonts}
\usepackage{algorithmic}
\usepackage{algorithm}
\usepackage{array}
\usepackage[caption=false,font=normalsize,labelfont=sf,textfont=sf]{subfig}
\usepackage{textcomp}
\usepackage{stfloats}
\usepackage{url}
\usepackage{verbatim}
\usepackage{graphicx}
\usepackage{cite}
\usepackage{balance} 
\IEEEoverridecommandlockouts
\usepackage{cite}
\usepackage{amsmath,amssymb,amsfonts,amsthm}
\usepackage{algorithmic}
\usepackage{graphicx}
\usepackage{textcomp}
\usepackage{algorithm,algorithmic}
\usepackage{xcolor}
\usepackage{mathtools}
\usepackage{bm}
\usepackage{hyperref}
\usepackage{placeins}
\usepackage{float}

\usepackage{graphics} % for pdf, bitmapped graphics files
\usepackage{epsfig} % for postscript graphics files
\usepackage{mathptmx} % assumes new font selection scheme installed
\usepackage{times} % assumes new font selection scheme installed
\usepackage{makecell}

\usepackage{color}
\usepackage{booktabs}

\DeclareMathAlphabet{\mathcal}{OMS}{cmsy}{m}{n}

\usepackage{float}
\def\BibTeX{{\rm B\kern-.05em{\sc i\kern-.025em b}\kern-.08em
    T\kern-.1667em\lower.7ex\hbox{E}\kern-.125emX}}

\newtheorem{remark}{Remark}

\newcommand{\bd}{\bm{d}}

\newcommand{\bg}{\bm{g}}

\newcommand{\bmm}{\bm{m}}

\newcommand{\bu}{\bm{u}}
\newcommand{\bv}{\bm{v}}

\newcommand{\bx}{\bm{x}}
\newcommand{\by}{\bm{y}}

\newcommand{\sN}{\mathcal{N}}

\newcommand{\sS}{\mathcal{S}}
\newcommand{\sT}{\mathcal{T}}

\newcommand{\sX}{\mathcal{X}}

\newcommand{\chen}[1]{\textcolor{red}{{[Chen: #1]}}}

\allowdisplaybreaks 
\begin{document}
\raggedbottom

\title{ Model-Free Coordinated Optimization of IBR Controllers for Enhanced Grid-Level Transient Dynamic Performance
}

\author{Haowen Xu,~\IEEEmembership{Graduate Student Member,~IEEE,}
        Xin Chen,~\IEEEmembership{Member,~IEEE}
        \thanks{ H. Xu and X. Chen are with the Department of Electrical and Computer Engineering, Texas A\&M University, USA. Emails: haowen\_xu@tamu.edu, xin\_chen@tamu.edu. \emph{(Corresponding author: Xin Chen)}.}
        \thanks{This work was supported in part by NSF AMPS 2523934 and in part by NSF CAREER 2541998. 
        }
}
\maketitle 

\begin{abstract}
With the increasing penetration of inverter-based resources (IBRs) in power grids, system-level coordinated optimization of IBR controllers has become increasingly important for maintaining overall system stability. Unlike most existing methods that rely on simplified or linearized dynamic models and focus on small-signal stability or isolated tuning of individual facilities, this paper proposes a novel simulation-based, model-free framework for the coordinated optimization of IBR controllers to enhance grid-level transient dynamic performance. The framework uses a high-fidelity power system simulator to accurately evaluate grid transient dynamic responses, and a projected multi-point zeroth-order optimization algorithm with adaptive moment estimation, termed PMZO-Adam, is proposed to solve the problem in a model-free manner, thus eliminating the need for explicit mathematical models of complex nonlinear system dynamics. The proposed framework enables direct optimization of grid transient dynamic behavior and  system-wide coordinated tuning of IBR controllers. Extensive simulations demonstrate the effectiveness of the proposed approach in optimizing IBR control parameters to improve grid transient frequency response under large disturbances.
\end{abstract}


\section{Introduction}\label{sec:introduction}

\IEEEPARstart{T}{he} electric power grid is undergoing a profound transformation driven by the increasing integration of inverter-based resources (IBRs), such as wind turbines, solar panels, and battery energy storage systems. Unlike conventional synchronous generators with large rotating masses \cite{gu2022power}, IBRs interface with the grid through power electronic inverters. Consequently, power systems with high IBR penetration exhibit fundamentally different dynamic behaviors, reduced system inertia, and increased challenges for maintaining grid stability \cite{milano2018foundations}. The dynamics of IBRs are primarily governed by their control strategies, which are broadly categorized as grid-forming (GFM) and grid-following (GFL) controls \cite{markovic2021understanding,gupta2025revisiting}. 
GFL inverters rely on an external grid voltage for synchronization, typically using a phase-locked loop (PLL) to estimate the grid voltage phase angle and frequency. In contrast, GFM inverters establish an internal voltage phasor and synchronize with the grid through their own voltage and frequency dynamics. 
As IBRs are interconnected through the power network, effective grid-level coordination of GFM and GFL controllers becomes essential for maintaining overall system stability and enhancing transient performance. 

%\chen{the key information of each paragraph should be clear and clean. Here, the first paragraph just needs to introduce the problem "coordination of IBR controls for grid stability". How to do that and literature review can be the next paragraph. Do not mix them.}

Conventionally, IBR control settings and parameters are determined empirically by the original equipment manufacturer or facility operator, based on operational experience and tuned independently at the device level \cite{Arzani2018StochasticSIA}.  
A number of studies have applied control-theoretic methods, such as eigenvalue analysis, root locus, and small-signal modeling \cite{Dagal2025EnhancingDCA, Chen2022OnPCA, Zhang2020AnIAA}, and model-based optimization approaches \cite{ArraoVargas2022ChallengesAMA, Khan2024GridformingCFA, Ducoin2024AnalyticalDOA}, to derive optimal controller parameter settings.
However, these model-based methods typically rely on simplified or linearized system dynamic models and primarily focus on small-signal stability, which may fail to capture the nonlinear dynamics of practical systems and can lead to suboptimal or even unstable performance under large disturbances \cite{Ma2019ChaosABA}. 

To overcome the limitations of model-based methods, data-driven techniques have been explored, including reinforcement learning \cite{Vlachogiannis2004ReinforcementLFA, Deng2024StochasticDPA} and meta-heuristic algorithms, such as particle swarm optimization and genetic algorithms \cite{Verdejo2020ImplementationOPA, Bano2024IntelligentCAA}. Nevertheless, these data-driven approaches generally lack performance guarantees, raising concerns about their optimality and reliability. 
Furthermore, many existing studies focus on optimizing the parameters of an individual IBR independently \cite{li2024zeroth, zalkind2020automatic}. Such localized facility-level control parameter tuning overlooks the strong dynamic coupling and interactions among multiple IBRs across the power network, which may adversely affect power system stability.

This paper focuses on the grid-level coordinated optimization of control parameters for both GFM and GFL IBRs to enhance overall system transient dynamic performance under large disturbances. Due to the complex nonlinear dynamics of power systems, transient stability analysis is primarily conducted using time-domain simulations in current industry practice. Mature power system simulation tools, such as PSSE \cite{psse}, PSCAD \cite{pscad}, and PowerWorld \cite{powerworld}, along with continuously advancing simulation technologies such as digital twins \cite{pan2020digital} and faster-than-real-time simulation \cite{8274433,10318675}, enable fast and scalable simulations for accurate stability assessment. Moreover, high-fidelity simulators are widely developed and utilized in the power industry for grid planning and operation, e.g., those embedded in energy management systems, which are readily available for use.

\begin{figure}
    \centering
    \includegraphics[width=1\linewidth]{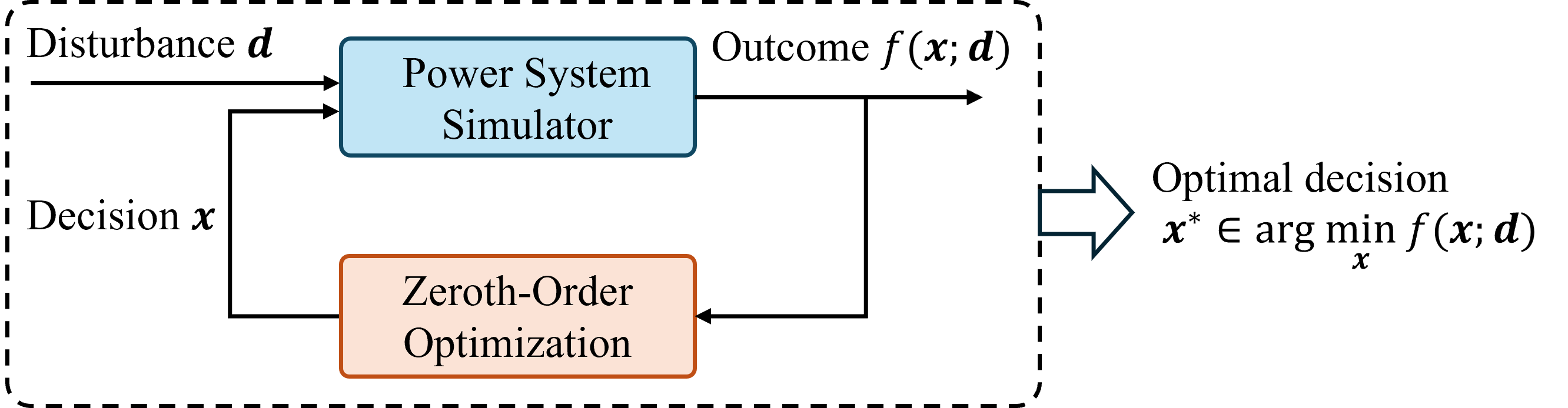}
    \caption{The simulation-based model-free grid optimization framework. %\chen{separate the input arrows of d and x, place them symmetrically} % (this iterative framework solves the optimization problem $\min_x f(\bx;\bd)$ to obtain the optimal decision $\bx^*$ without requiring an explicit mathematical model of $f(\bx;\bd)$). %\chen{[\checkmark] change "=" to "$\in$" in case the optimal solution is not unique}%\chen{only two lines for "Zeroth..Algorithm", adjust the figure}\chen{Figure caption can provide more information; add $min_x f(x;w)$; change the variable notations to be consistent; Avoid any abbreviations in figure, use zeroth-order optimization instead so that no need to check what ZO mean.}\chen{For all files/figure names, do not use space, which may cause format issue; you can use underscore "\_" instead.}
    }
    \label{fig:structure}
\end{figure}

Motivated by these capabilities, this paper proposes a novel framework that integrates a \emph{power system simulator} with \emph{zeroth-order optimization (ZO) algorithms} \cite{liu2020primer, nesterov2017random} to optimize IBR control parameters in a model-free manner, thus eliminating the need for explicit mathematical models of complex nonlinear power system dynamics. As illustrated in Figure~\ref{fig:structure}, given a decision $\bx$ of IBR control parameters and under disturbances $\bd$, 
the power system simulator acts as a black-box oracle that can accurately evaluate transient dynamic outcome $f(\bx;\bd)$, such as frequency nadir/zenith, the rate of change of frequency (RoCoF), and the maximum phase angle \cite{tuo2023convolutional,gan2000stability}. ZO algorithms \cite{liu2020primer, nesterov2017random}, also known as gradient-free methods, offer an effective approach
for solving black-box optimization problems using only zeroth-order information, i.e., function evaluations or simulation outputs. Essentially, ZO methods mimic gradient-descent-type algorithms but replace the true gradient with a zeroth-order gradient estimator constructed from perturbed function evaluations; see Section~\ref{sec:zo} for a detailed introduction. As a result, ZO methods provide strong theoretical convergence guarantees \cite{nesterov2017random,chen2025continuous}
and have been used in a broad range of applications, including reinforcement learning \cite{li2019distributed},  neural network training \cite{chen2017zoo}, power system control \cite{chen2021safe}, etc.
%sensor management \cite{liu2020primer}, etc. 

 \emph{Contributions}. This paper proposes a simulation-based, model-free framework for coordinated optimization of IBR control parameters to enhance grid-level \emph{transient} dynamic performance, including mitigating frequency nadir/zenith deviations and suppressing frequency oscillations. Building on existing adaptive ZO methods \cite{chen2019zo}, we develop a \emph{projected multi-point ZO algorithm with adaptive moment estimation (PMZO-Adam)} and integrate it with high-fidelity power system simulations. The main contribution lies in this integrated algorithmic framework and its application to jointly tuning dynamically coupled GFM and GFL controllers under large disturbance scenarios. This problem is particularly suited to simulation-based zeroth-order optimization, as simulations capture complex controller interactions and nonlinear transient responses, for which analytical objective gradients are difficult to obtain. 
 The key contributions and features of the proposed framework are summarized below: 
 \begin{itemize}
     \item[1)] \emph{Integrated algorithm}. The proposed PMZO-Adam algorithm integrates parameter projection, multi-point ZO gradient estimation, and adaptive moment updates into a unified parallel simulation workflow for coordinated IBR controller tuning. Projection enforces prescribed bounds on the updated controller parameters, while averaging gradient estimates over multiple perturbation directions reduces estimation variance. Parallel simulations across perturbations and disturbance scenarios reduce evaluation time, and Adam improves optimization efficiency through momentum and coordinate-wise adaptive step sizes \cite{kingma2014adam}. 
     \item[2)] \emph{Direct optimization of grid transient performance}. The framework jointly optimizes GFM and GFL controller parameters, including inertia, damping, PLL, and inner-loop parameters, 
     leveraging high-fidelity simulations to directly optimize grid transient performance. It captures interactions among network-connected IBRs and balances system responses across multiple disturbance scenarios, extending the optimization scope beyond individual-device performance and small-signal criteria.

\item[3)] \emph{Model-free interface and flexible formulation}. The optimizer uses only simulation-based objective evaluations, requiring no explicit analytical models of complex system dynamics. Here, \emph{model-free} refers to the optimization interface, while transient performance is evaluated through power system simulators. This \emph{flexible} framework  accommodates diverse power network configurations, disturbances, and performance metrics, and extends to many other decisions, such as IBR power references for dynamics-aware power dispatch. 

\end{itemize}

In addition, a high-fidelity electromagnetic transient (EMT) model of a modified IEEE 39-bus system with ten IBRs is developed in Matlab Simulink as the simulation platform. Extensive simulations demonstrate the effectiveness of the proposed framework in improving grid transient responses and characterize its optimization performance.

Moreover,
the proposed framework is intended for offline application, with IBR controller parameters optimized through simulations for subsequent deployment. This setting draws on available power system simulation models and aligns with industry practice, where controller parameters are typically configured in advance and adjusted infrequently.

% Additionally, the proposed framework is intended for offline application, where key IBR control parameters are optimized using simulations and subsequently deployed in real-world systems. This setting leverages the practical availability of accurate power system simulators and is consistent with the industry practice, in which facility control parameters are typically predetermined and adjusted infrequently.

The remainder of this paper is organized as follows: Section \ref{sec:controller} introduces the control architectures and parameters of GFL and GFM inverters. Section \ref{sec:algorithm} develops the model-free optimization framework based on ZO algorithms. Section \ref{sec:simulation} presents experimental validation and performance analysis, and Section \ref{sec:conclusion} concludes the paper.

\section{Control Architectures and Parameters of GFL and \text{GFM} Inverters}\label{sec:controller}

% \chen{\text{GFM}s? GFLs?} 
% \chen{a short paragraph introducing what you are going to present.} 
% \chen{think of the function of a section. This section is to briefly introduce the control architectures and also control parameter. two goals: 1. introduce the control parameters which are the decision variables, 2. show how complex it is.}

% \chen{what's the different between inverter and converter? Statements need to be accurate and correct, e.g., IBR operating as either GFL or \text{GFM} converters, how about hybrid GFL/\text{GFM}, use words like typically, generally, primarily}

% \chen{try not to highlight "model", people might get confused as your approach is claimed model-free}

\begin{figure}
    \centering
    \includegraphics[width=1\linewidth]{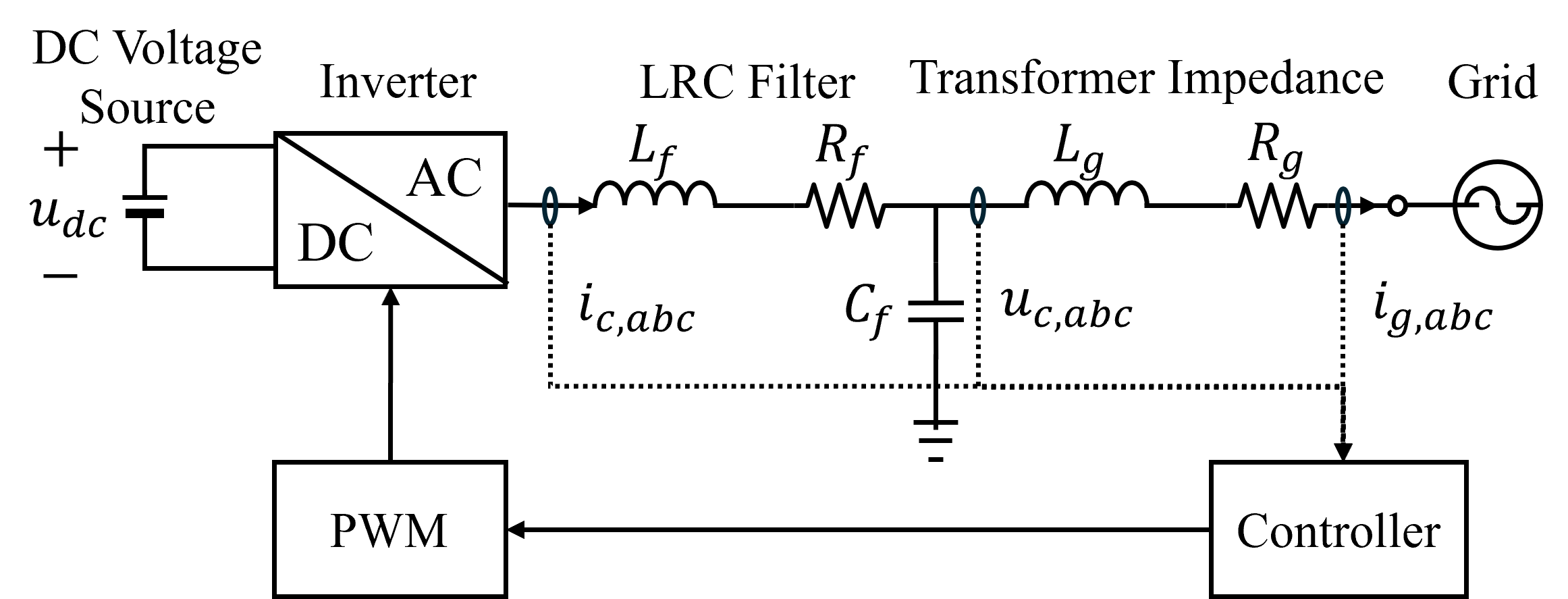}
    \caption{Typical configuration of a three-phase inverter connected to the power grid (the inverter is connected to the grid through a LRC filter with inductance $L_f$, resistance, $R_f$, and capacitance $C_f$ and a transformer modeled with inductance $L_g$ and resistance $R_g$).%\chen{[\checkmark] change "invertor" to "inverter"} %\chen{explain the parameters, like the inverter connect to the grid through a LRC filter with inductance $L_f$, resistance, $R_f, and capacitor C_f$ and a transformer modeled with $L_g, R_g$...} \chen{No need signal generator block. add some text notes like DC voltage source, inverter, change the sign of grounded, the dashed lines are not clear, add a line head }
    }
    \label{fig:1}
\end{figure}

This section presents the typical control architectures of \text{GFM} and GFL inverters and identifies the key control parameters that can be tuned to enhance grid dynamic performance. Figure \ref{fig:1} illustrates a typical configuration of a three-phase inverter system connected to the grid. The controller block processes measured signals, such as the three-phase inverter output current \(i_{c,abc}\), capacitor voltage \(u_{c,abc}\), and grid-side current \(i_{g,abc}\), and generates pulse-width modulation (PWM) switching signals for the power electronic switches. The GFL and \text{GFM} control strategies are further introduced below.

% Various IBRs utilize converters to interface with the grid, operating as either GFL or \text{GFM} converters \cite{gao2024stability, Wang2024DistributedCOA}. GFL converters function as controlled current sources, whereas \text{GFM} converters act as controlled voltage sources \cite{Wang2024DistributedCOA}.

% Figure \ref{fig:1} illustrates the configuration of a three-phase grid-connected converter system. This system employs a hierarchical control architecture that processes sensed signals, such as the three-phase converter output current \(i^{abc}_c\), capacitor voltage \(u^{abc}_c\), and line current \(i_{g}^{abc}\), through multiple control loops. These loops generate gate-drive signals for the power electronic switches using pulse-width modulation (PWM). The following sections describe the distinct control strategy models for GFL and \text{GFM} converters.

% \subsection{Control Model of GFLs}

\subsection{Grid-Following (GFL) Inverter Control}

% \chen{To type double quotes, use ``..." in latex not "..." }

% \chen{polish the introduction to be clear and accurate. You can ask GPT to generate a concise high-level introduction of GFL control, and adapt from it.} 

% \chen{
% Following the control process, the key components can be introduced firstly PLL, then droop loop, and current control loop in three separate paragraphs. }

% \chen{since you only gives a brief introduction, you can add sentences like "see [xx] [xx] for a more detailed introduction" to end.}
% \chen{at the end, summarize the key control parameters for optimization. These two also apply to the \text{GFM} control.}

\begin{figure}
    \centering
    \includegraphics[width=1\linewidth]{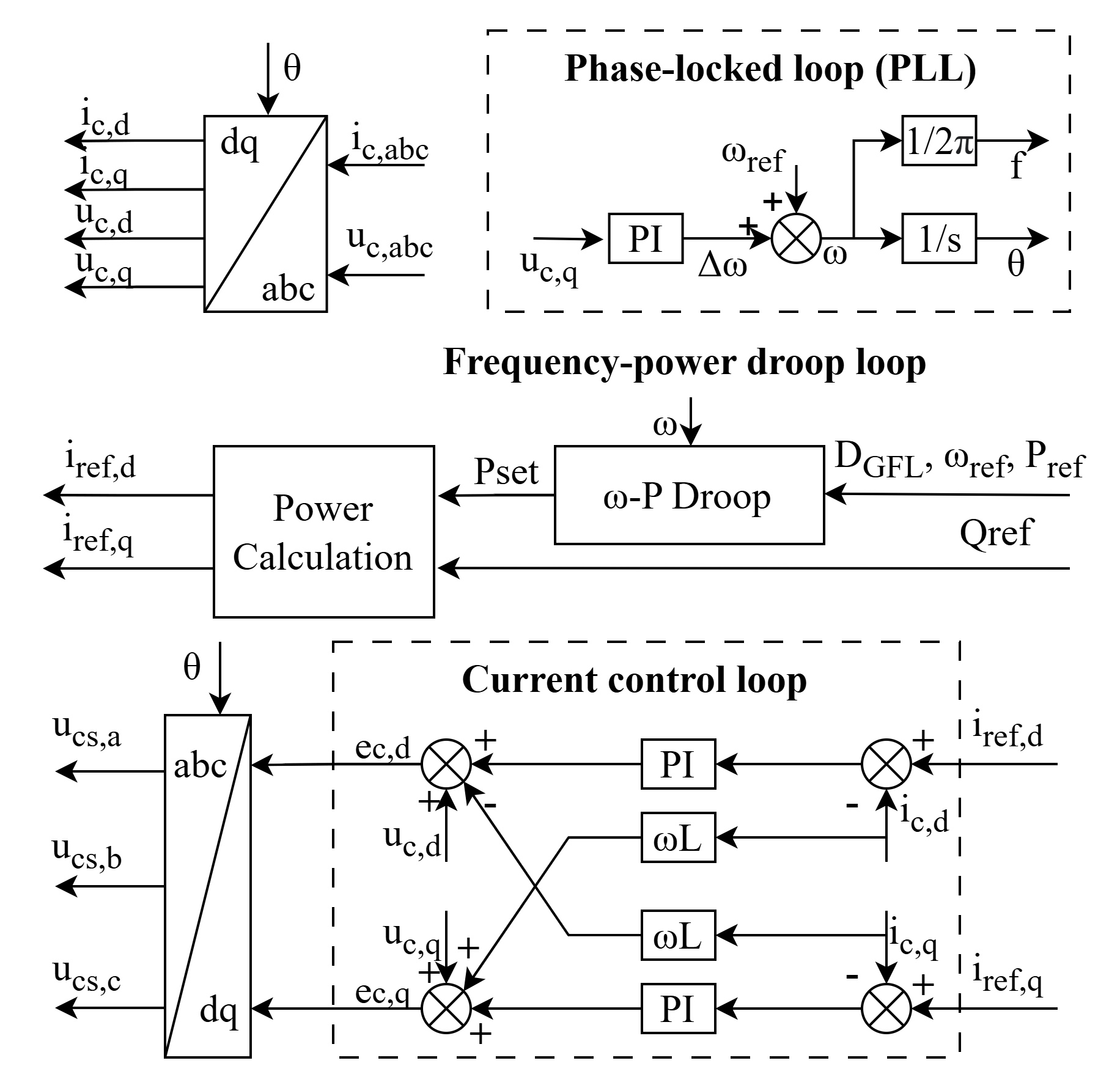}
    \caption{The phase-locked loop, frequency-power droop control, and current control loop in a GFL inverter (refer to \cite{Ducoin2024AnalyticalDOA} for more details). }% (refer to \cite{nestor2025data} for the function of each loop in detail).
    % \chen{change the caption to be The phase-locked loop, xxx, xxx in GFL inverter control.} \chen{explain each control blocks, add a title for each; rearrange them in the order or introduction }\chen{Here the figures are complex but only a simple caption. What you can do is use (...) to add a bit more explanations or directly cite references.}}
    \label{fig:2}
\end{figure}

GFL inverters are a widely deployed class of inverters that measure grid states to regulate their output current $i_{c,abc}$ and synchronize with the grid voltage phase angle and frequency \cite{du2020modeling}. As shown in Figure \ref{fig:2}, the three key control components of a GFL inverter include the PLL, the frequency-power droop loop, and the current control loop.

The PLL of GFL control is used to estimate the voltage phase angle and grid frequency \cite{Ducoin2024AnalyticalDOA}. Specifically, the PLL employs a proportional-integral (PI) controller to eliminate the frequency estimation error $\Delta \omega$ by driving 
 the q-axis component of the interconnection point voltage, $u_{c,q}$, to zero.
 The estimated frequency is obtained by adding $\Delta \omega$ to the nominal grid frequency $\omega_0$, and integrating this frequency yields the estimated phase angle $\theta$. Thus, the PLL control dynamic equations are formulated as \eqref{eq:pll}:
\begin{gather}
\left\{\begin{array}{l}
        \dot\phi_{\text{PLL}}=K_{i,\text{PLL}}\cdot(u_{c,q}-0)\\
        \Delta \omega=K_{p,\text{PLL}}\cdot(u_{c,q}-0)+\phi_{\text{PLL}}\\
        \dot \theta = \Delta \omega + \omega_0
        \end{array}\right.
\label{eq:pll}   
\end{gather}
where $K_{i,\text{PLL}}$ and $K_{p,\text{PLL}}$ are the integral and proportional PI parameters of PLL. $\phi_{\text{PLL}}$ is the PLL integrator state variable.

Frequency-power droop control is commonly incorporated into GFL inverter control to regulate active power output in response to variations in grid frequency \cite{ochoa2025optimal}. 
Based on the difference between a reference frequency $\omega_{\text{ref}}$ and the PLL's estimated frequency $\omega$, this droop control generates the active power setpoint $P_{\text{set}}$ according to \eqref{eq:fPdroop}, which is then used to determine current reference values ($i_{\text{ref},d}, i_{\text{ref},q}$).
% measures the estimated frequency $\omega$ from the PLL with a reference frequency $\omega_{ref}$. Based on this frequency error, the droop loop generates a power reference $P_{set}$, which is used to determine current reference values ($i_{ref,d}, i_{ref,q}$). The relationship is described by:
\begin{gather}
    P_{\text{set}}=D_{\text{GFL}}\cdot(\omega_{\text{ref}}-\omega)+P_{\text{ref}}.
\label{eq:fPdroop}   
\end{gather}
Here, $D_{\text{GFL}}$ denotes the droop damping coefficient that influences the frequency response and power-sharing characteristics, and $P_{\text{ref}}$ is the predefined active power reference.
% \chen{the subscript format for text in notations is not consistent. You should use mathrm, like $D_\mathrm{GFL}$ and $P_\mathrm{ref}, P_{\mathrm{set}}$. Please correct all the notations and use this format in the future.}

In addition, the current control loop generates three-phase PWM control signals ($u_{cs,a}, u_{cs,b}, u_{cs,c}$) to modulate inverter switching and produce the output current  $i_{c,abc}$. This loop includes two PI controllers with tunable control parameters, $K_{p,i,\text{GFL}}$ and $K_{i,i,\text{GFL}}$, which significantly influence the transient response of the GFL inverter.

\emph{Key GFL Control Parameters}.
As a result, the dynamic behavior of GFL inverters is
governed primarily by the control loops described above and is highly sensitive to the tuning of several key parameters, including the PLL gains $K_{p,\text{PLL}}, K_{i,\text{PLL}}$, the droop control coefficient $D_{\text{GFL}}$, and the current control gains $K_{p,i,\text{GFL}}, K_{i,i,\text{GFL}}$. Coordinated optimization of these parameters is essential to ensure stable operation under varying grid conditions. %, as addressed in the subsequent section.  %This issue is examined in the subsequent sections of this study.

% The inverter's transient performance is highly sensitive to the tuning of several key parameters in these control loops, including the PLL gains ($K_{p,PLL}, K_{i,PLL}$), the droop coefficient ($D_{GFL}$), and the current loop gains ($K_{p,i,GFL}, K_{i,i,GFL}$). Optimizing these parameters is therefore essential for ensuring stable operation under varying grid conditions, a task addressed in the subsequent sections of this study.

%\addtolength{\textheight}{-3cm}   % This command serves to balance the column lengths
                                  % on the last page of the document manually. It shortens
                                  % the textheight of the last page by a suitable amount.
                                  % This command does not take effect until the next page
                                  % so it should come on the page before the last. Make
                                  % sure that you do not shorten the textheight too much.

%%%%%%%%%%%%%%%%%%%%%%%%%%%%%%%%%%%%%%%%%%%%%%%%%%%%%%%%%%%%%%%%%%%%%%%%%%%%%%%%

% \subsection{Control Model of \text{GFM}s}

\subsection{Grid-Forming (\text{GFM}) Inverter Control}

% \chen{You can start this subsection with "In contrast to the GFL control, \text{GFM} control ..."}

\begin{figure}
    \centering
    \includegraphics[width=1\linewidth]{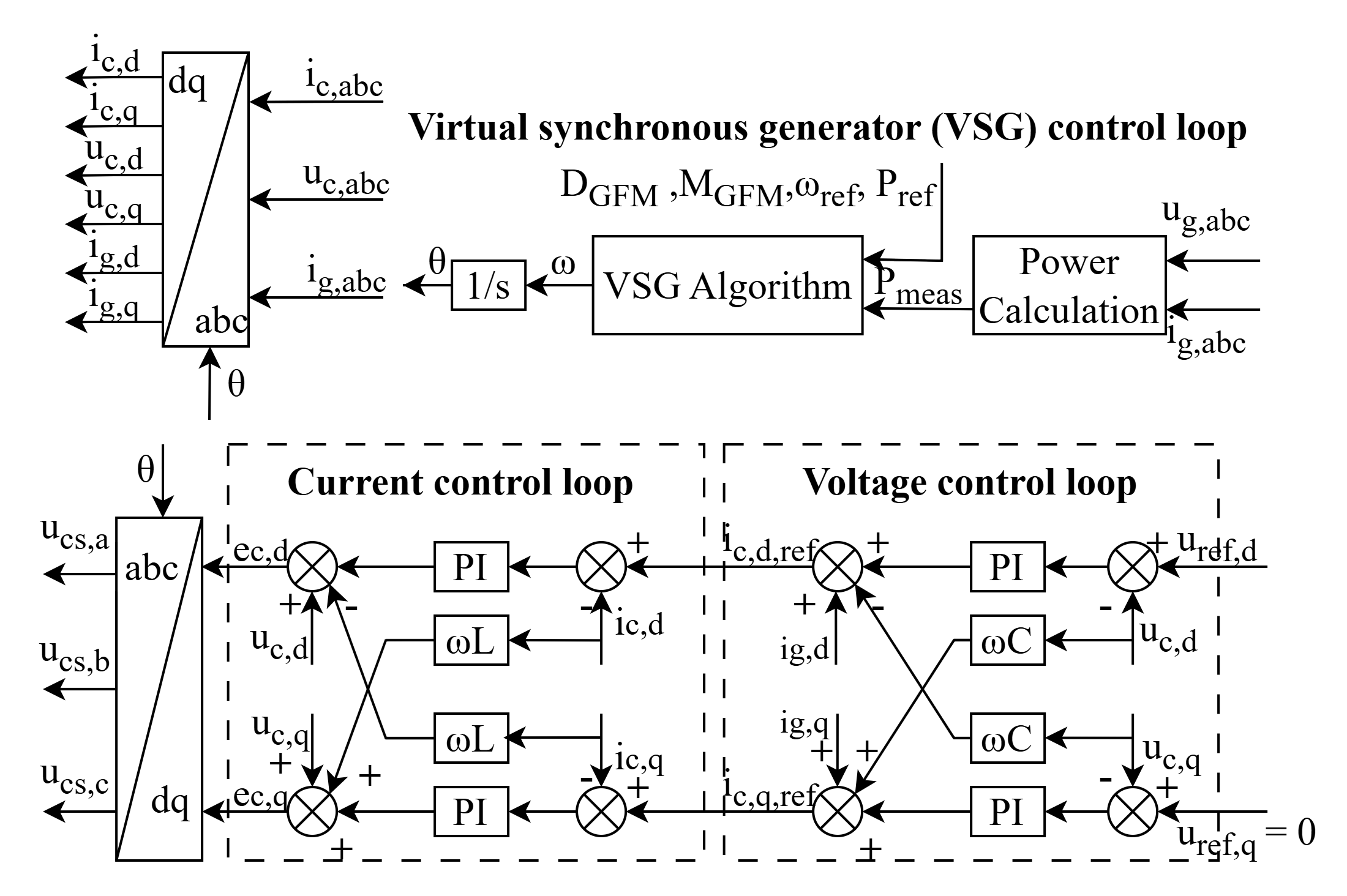}
    \caption{The virtual synchronous generator (VSG) control loop, voltage control loop, and current control loop in a \text{GFM} inverter (refer to \cite{Ducoin2024AnalyticalDOA} for more details). 
    % \chen{make similar changes as the GFL figure}
    }
    \label{fig:3}
\end{figure}

In contrast to GFL control, \text{GFM} control has the capability to form and regulate the inverter's frequency and voltage \cite{rathnayake2021grid}. Typical \text{GFM} control strategies include droop-based control, virtual synchronous generator (VSG) control, and virtual oscillator control (VOC) \cite{tozak2024modeling}, etc. In this paper, we adopt the VSG control as a representative example for \text{GFM} control, and its architecture is illustrated in Figure \ref{fig:3}. This scheme consists of three main components: the VSG control loop, the voltage control loop, and the current control loop.

% the ability to form the frequency and voltage of the system by emulating the control characteristics of synchronous generators\cite{rathnayake2021grid}. Various \text{GFM} control strategies exist, such as droop control, virtual synchronous generator (VSG) control, and virtual oscillator control (VOC) \cite{tozak2024modeling}, etc. The VSG control is adopted in this paper as an example, with its architecture depicted in Figure \ref{fig:3}. The control scheme includes three main components: the VSG algorithm, the voltage control loop, and the current control loop.

% \chen{This paper does not necessarily have to adopt the VSG, you can just say, for example, the VSG control ...}

The VSG control loop consists of two core components: a power-frequency controller and an excitation controller. The power-frequency control dynamics is given by \eqref{eq:vsg} \cite{tozak2024modeling}:
\begin{gather}
    M_{\text{GFM}}\cdot\dot \omega= ({P_{\text{ref}}-P_{\text{meas}}})/{\omega} - D_{\text{GFM}}\cdot(\omega-\omega_{\text{ref}})
\label{eq:vsg}   
\end{gather}
where $M_{\text{GFM}}$ denotes the emulated inertia parameter, $D_{\text{GFM}}$ is the damping coefficient, $\omega_{\text{ref}}$ is the frequency reference, $\omega$ is the frequency established by the \text{GFM} inverter, $P_{\text{ref}}$ is the active power reference, and $P_{\text{meas}}$ is the measured active power, which is typically filtered.
Equation \eqref{eq:vsg} emulates the swing dynamics of a synchronous generator, thus enabling the \text{GFM} inverter to provide synthetic inertia and damping support to the grid.
%\chen{Is equation \eqref{eq:vsg} correct for VSG?  add a reference}

% is the momentum of inertia, $D_{\text{GFM}}$ is the damping coefficient,
% $\omega_{ref}$ is the frequency reference, $\omega$ is the formed frequency by the \text{GFM} inverter, $P_{ref}$ is the active power reference, and $P_{meas}$ is the measured active power, typically filtered. This equation mimics the swing equation of a synchronous generator, enabling the \text{GFM} inverter to deliver synthetic inertia and damping to the grid.

As illustrated in Figure \ref{fig:3},
the voltage control loop regulates the inverter output voltage $u_{c,abc}$ to track the voltage reference ($u_{\text{ref},d},u_{\text{ref},q}=0$) and generates current references ($i_{c,d,\text{ref}},i_{c,q,\text{ref}}$) for the current control loop. These current references are then converted into PWM signals, following a structure similar to that of GFL inverters.

\emph{Key \text{GFM} Control Parameters}.
In summary, the performance of a \text{GFM} inverter to provide inertia and damping support depends on the coordinated interaction of these control loops described above. Particularly, the dynamic response of the inverter system is highly sensitive to several key parameters, including the inertia parameter $M_{\text{GFM}}$, the damping coefficient $D_{\text{GFM}}$, and the voltage/current loop gains ($K_{p,v,\text{GFM}}, K_{i,v,\text{GFM}}, K_{p,i,\text{GFM}}, K_{i,i,\text{GFM}}$). Appropriate tuning of these parameters is therefore critical for maintaining stable operation and achieving the desired dynamic performance, which is addressed in the following section.

% the \text{GFM} inverter's ability to provide inertia and frequency support depends on the interaction between the VSG algorithm and the inner control loops. The system's dynamic response is highly sensitive to the inertia constant ($M_{\text{GFM}}$), the damping coefficient ($D_{\text{GFM}}$), and the voltage/current loop gains ($K_{p,v,\text{GFM}}, K_{i,v,\text{GFM}}, K_{p,i,\text{GFM}}, K_{i,i,\text{GFM}}$). Optimizing these parameters is critical for maintaining stability, which will also be the focus of the following sections.

\section{Problem Formulation and Algorithm Design}\label{sec:algorithm}

% \chen{So far, we haven't introduced what problem formulation is. People will get confused what do you really want to address.
% So it would be better to change this section structure to be A. Problem Formulation to introduce the problem. B. ZO algorithms and ZO-AdaMM. C. Your algorithm}

% \chen{This introduction paragraph needs to be concise and high-level, just summarize what are presented. no need detailed explanations}

In this section, we first formulate the coordinated optimization problem for IBR control parameters to enhance transient grid dynamic performance. We then introduce the preliminaries of ZO and the Adam method. Lastly, we propose the PMZO-Adam algorithm for simulation-based model-free optimization to solve the formulated problem.

\subsection{Problem Formulation} \label{sec:P_F}

Consider a generic problem formulation \eqref{eq:base_obj} for grid-level coordinated optimization of IBR control parameters:
% As we mentioned in Introduction, we plan to solve the optimization problem $\min_x f(\bx;\bd)$. Moreover, we can have a more general objective formulation as:
\begin{align}
    \min_{\bx\in\sX} \ f(\bx) \coloneqq \sum_{s\in\sS} \alpha_s\cdot \hat{f}(\bx;\bd_s),
    \label{eq:base_obj}
\end{align}
where the decision variable $\bx$ denotes the key control parameters of all IBRs in the grid, and $\sX$ is the feasible set, such as the physical upper and lower limits of $\bx$. $s$ denotes a disturbance scenario in the scenario set $\sS$, and $\bd_s$ represents a specific disturbance, e.g., load change and line contingency. $\alpha_s$ is the weight factor of scenario $s$, and $\hat{f}$ represents the grid transient dynamic performance metrics to be optimized, such as the frequency nadir or zenith, RoCoF, frequency fluctuations, maximum phase angle, etc. Hence, problem \eqref{eq:base_obj} aims to optimize the overall grid dynamic performance, weighted across a predefined set of disturbance scenarios.

% $\sX$ is the feasible set, like the upper and lower bounds, $s$ denotes a disturbance scenario in the scenario set $\sS$, and $\bd_s$ is the disturbance, such as load change and line contingency, $\alpha_s$ is the weight. $\hat{f}$ represents the grid transient dynamic performance metrics to be optimized, such as the frequency nadir/zenith and frequency fluctuations. 

Specific grid transient dynamic performance metrics $\hat{f}$ can be flexibly adapted based on practical considerations.
For instance, we aim to mitigate frequency nadir or zenith and suppress frequency oscillations after a disturbance. Accordingly, $\hat{f}$ can be formulated as \eqref{eq:objective}, where the scenario index $s$ is omitted for notational simplicity. 
% we consider one disturbance scenario for one time, and focus on minimizing frequency deviations and suppressing oscillations. The objective function is formulated to evaluate the time-domain dynamic response of the system:
\begin{align} \label{eq:objective}
    \hat{f}\coloneqq \sum_{i\in\mathcal{N}_{\mathrm{opt}}} \Big[ &\lambda\cdot \max_{t \in \mathcal{T}_{\mathrm{dev}}} |\omega_{i,t} - \omega_{0}| \nonumber \\ 
    & + (1-\lambda) \cdot  \frac{1}{|\mathcal{T}_{\mathrm{osc}}|} \sum_{t \in \mathcal{T}_{\mathrm{osc}}} (\omega_{i,t} - \bar{\omega}_{i,\sT_{\mathrm{osc}}})^2   \Big].
\end{align}
%\chen{change the format of opt, dev and osc, etc., that have text meanings, using mathrm, as well as all other places they appear}
Here, $i$ denotes the bus index, and $\mathcal{N}_{\mathrm{opt}}$ is the set of buses whose frequencies are considered in the optimization. $t$ denotes the time index. $\sT_{\mathrm{dev}}\coloneqq \{t_\mathrm{d}, t_\mathrm{d}+\Delta t,\cdots, T\}$
denotes the discrete time window over which the maximum frequency deviation is evaluated, where $t_\mathrm{d}$ is the disturbance occurrence time, $\Delta t$ is the time step size, and $T$ is the final time step. $\omega_{i,t}$ is the frequency of bus $i$ at time $t$, and $\omega_{0}$ is the nominal frequency (e.g., 60 Hz in the U.S.). $\sT_{\mathrm{osc}}\coloneqq \{t_\mathrm{o}, t_\mathrm{o}+\Delta t,\cdots, T\}$
denotes the discrete time window over which the post-disturbance frequency oscillation is evaluated, where $t_\mathrm{o}$ is the corresponding initial time step. $\bar{\omega}_{i,\sT_{\mathrm{osc}}}$ denotes the average value of the frequency $\omega_{i,t}$ over the time window $\sT_{\mathrm{osc}}$. 

Hence, the grid dynamic performance metric $\hat{f}$ defined in \eqref{eq:objective} is a weighted multi-objective function that balances the maximum frequency deviation (first term) and the empirical variance of the frequency (second term), which captures post-disturbance oscillations, with the weight factor $\lambda \in [0,1]$.

The decision variable vector $\bx$ consists of the key control parameters of IBRs. As introduced in Section \ref{sec:controller}, $\bx$ includes $\{K_{p,\text{PLL}}, K_{i,\text{PLL}}, D_{\text{GFL}}, K_{p,i,\text{GFL}}, K_{i,i,\text{GFL}}\}$ for a GFL inverter, and $\{M_{\text{GFM}}, D_{\text{GFM}}, K_{p,v,\text{GFM}}, K_{i,v,\text{GFM}}, K_{p,i,\text{GFM}}, K_{i,i,\text{GFM}}\}$ for a GFM inverter. Each entry $x_j$ in $\bx$ is subject to the feasible interval $[\underline{x}_j, \overline{x}_j]$, and thus the feasible set is $\mathcal{X} \coloneqq \prod_j \,[\underline{x}_j, \overline{x}_j]$.

\begin{remark}
    \normalfont
A critical challenge in solving \eqref{eq:base_obj} is the absence of an explicit mathematical model of the grid transient dynamic performance function $\hat{f}(\bx;\bd_s)$. This function represents the input-output mapping governed by the complex nonlinear power system dynamics and is generally intractable to derive. To address this challenge, we propose integrating a high-fidelity power system simulator with ZO algorithms to solve \eqref{eq:base_obj} in a model-free manner. In this framework, the simulator acts as a black-box oracle that outputs the value of $\hat{f}(\bx;\bd_s)$ for a given decision $\bx$ and disturbance $\bd_s$.
 As our framework enables simulation-based model-free optimization, it can flexibly accommodate any transient dynamic performance metrics $\hat{f}$ constructed from simulation outputs.  \qed
\end{remark}

% Each individual parameter $x_j$, where $j$ denotes the index of the specific controller gain or physical constant (such as virtual inertia or damping), is constrained within a physically realizable range $[\underline{x_j}, \overline{x_j}]$. Consequently, the feasible set is defined as the Cartesian product of these ranges, $\mathcal{X} = \prod_j [\underline{x_j}, \overline{x_j}]$. 

\subsection{Zeroth-Order Optimization and Adam Method} \label{sec:zo}

This subsection presents the preliminaries of ZO algorithms and the Adam method, which incorporates  adaptive moment estimation to accelerate convergence.

% the basic ZO algorithm, specifically two-point zeroth-order algorithm. And then we adopt the AdaMM algorithm into the zeroth-order framework to get better converge performance.
\subsubsection{Zeroth-Order Optimization (ZO)}

ZO methods have been widely applied to solve black-box optimization and control problems. In essence, ZO methods mimic gradient-based algorithms but replace the unavailable gradient with a zeroth-order gradient estimator constructed from perturbed function evaluations. Depending on the number of queried function evaluations at each iteration,  ZO methods can be categorized into three classes: single-point, two-point, and multi-point schemes \cite{chen2022improve,chen2026regression,liu2020primer}. 

Consider solving the unconstrained version of \eqref{eq:base_obj} without the feasible set $\sX$. The two-point ZO method is given by: 
\begin{gather}
  \quad  \bx_{k+1}=\bx_{k}-\eta\cdot G_f^{(2)}(\bx_k;r,\bu_k), \quad k = 0,1,\cdots,
\label{eq:TZO-Main}
\end{gather}
where $\eta$ denotes the step size, and $G_f^{(2)}(\bx;r,\bu)$ is the two-point ZO gradient estimator defined as \eqref{eq:TZO-Gra}:
\begin{gather}
    G_f^{(2)}(\bx;r,\bu):=\frac{d}{2r}\Big(f(\bx+ r \bu)-f(\bx- r \bu)\Big)\bu. 
\label{eq:TZO-Gra}
\end{gather}
Here, $d$ denotes the dimension of $\bx$, $r\!>\!0$ is a scalar parameter called the smoothing radius, and $\bu \in\mathbb{R}^d$ is a random direction vector independently sampled at each iteration $k$ from the uniform distribution on the unit sphere $\mathbb{S}_{d-1}$. Two function evaluations at the points $\bx\!+\! r \bu$ and $\bx\!-\! r\bu$ are used to construct $G_f^{(2)}(\bx;r,\bu)$, which serves as a biased estimator of the true gradient $\nabla f(\bx)$; see \cite{nesterov2017random} for more details.

\subsubsection{The Adam Method} 

The adaptive moment estimation (Adam)  \cite{kingma2014adam} method is one of the most widely used first-order gradient-based optimization methods in machine learning. It leverages past gradients to update both descent directions and step sizes, which effectively accelerates convergence. Reference \cite{chen2019zo} extends this idea to the black-box optimization settings, where explicit gradients are unavailable. %, and proposes the ZO-AdaMM algorithm. 
Specifically, the Adam update at each iteration $k$ is given by \eqref{eq:ZOAdaMM}:
\begin{subequations} \label{eq:ZOAdaMM}
    \begin{align}
    &\bmm_k = \beta_1 \bmm_{k-1} + (1 \!-\! \beta_1)  {\bg}_k, \  \bv_k = \beta_2 \bv_{k-1} + (1 \!-\! \beta_2)  {\bg}_k^2,  \label{eq:ZOAdaMM:m}\\
    &\hat{\bmm}_k = \bmm_k / (1 - \beta_{1}^k),\quad   \hat{\bv}_k = \bv_k / (1 - \beta_{2}^k), \label{eq:ZOAdaMM:up}\\
    &\bx_{k+1} = \bx_k - \eta \frac{\hat{\bmm}_k}{\sqrt{\hat{\bv}_k} + \epsilon}. \label{eq:ZOAdaMM:x}
\end{align}
\end{subequations}
In \eqref{eq:ZOAdaMM:m}, $\bg_k$ denotes the gradient estimate at iteration $k$. $\bm{m}_k$ and $\bv_k$ represent the first and second moment estimates, respectively. %The parameters \(\beta_1, \beta_2 \in [0,1)\) are decay coefficients.
Equation \eqref{eq:ZOAdaMM:m} updates the exponential moving averages of the gradient ($\bm{m}_k$) and the squared gradient ($\bv_k$), where the hyper-parameters \(\beta_1, \beta_2 \in [0,1)\) control the exponential decay rates of these moving
averages. 
Equation \eqref{eq:ZOAdaMM:up} computes the  bias-corrected first and second moment estimates, where $\beta_{1}^k$ and $\beta_{2}^k$ denote $\beta_{1}$ and $\beta_{2}$ to the power of $k$, respectively. In \eqref{eq:ZOAdaMM:x}, $\epsilon$ is a very small positive constant introduced to prevent division by zero. See \cite{kingma2014adam} for more details on the algorithm.

\subsection{Our Algorithm for Coordinated IBR Control Optimization}

Built upon the methods described above, we propose the PMZO-Adam algorithm (i.e., Algorithm \ref{alg:PMZO-AdaMM_Unified}) to effectively solve the coordinated IBR control parameter optimization problem \eqref{eq:base_obj}. To further reduce the variance in gradient estimation and improve convergence, we adopt a multi-point ZO method that performs $N$ batches of two-point evaluations and uses their average as the gradient estimate. 

Specifically, at each iteration $k$, Algorithm \ref{alg:PMZO-AdaMM_Unified} 
samples \(N\) random perturbation directions \(\{{\bu}_{k,n}\}_{n=1}^N\) from the uniform distribution on the unit sphere \(\mathbb{S}_{d-1}\) (Step \ref{step:sample}).
Then, high-fidelity power system simulations are executed in parallel to evaluate the grid transient dynamic performance $\hat{f}$ for the perturbed decisions $\bx_k \!+\! r_k \bu_{k,n}$ and $\bx_k \!-\! r_k \bu_{k,n}$ under each disturbance $\bd_s$ for $s \in \sS$. This yields the objective function evaluations $f(\bx_k \!+\! r_k \bu_{k,n})$ and $f(\bx_k \!-\! r_k \bu_{k,n})$ in \eqref{eq:base_obj} for $n = 1, \ldots, N$ (Step \ref{step:simu}). Next, these function evaluations are used to construct the multi-point ZO gradient estimate $\bg_k$ as defined in \eqref{eq:multiZO}, which averages the two-point gradient estimates  \eqref{eq:TZO-Gra} over $N$ batches to reduce estimation variance. 
In Step \ref{step:update}, the Adam method \eqref{eq:ZOAdaMM} is employed to update the iterate.
Moreover, the projection operator, $\text{Proj}_{\sX}(\by)\coloneqq \arg \min_{\bx\in \sX}\|\bx-\by\|^2_2$, is incorporated into \eqref{eq:iterate} to ensure that all iterates remain within the feasible set $\sX$. In addition, 
the stepsize $\eta_k$ and smoothing radius $\delta_k$ decay monotonically according to \eqref{eq:szsr} in Step \ref{step:szsr} to enhance  convergence, enabling robust exploration in the early iterations while allowing precise fine-tuning near convergence.

% we propose the PMZO-AdaMM algorithm. This method introduces multi-point gradient averaging and projection to the basic ZO-AdaMM framework to handle high variance and enforce physical constraints.

% Algorithm \ref{alg:PMZO-AdaMM_Unified} describes the complete closed-loop optimization framework. During each iteration \(k\), the algorithm uniformly samples \(N\) random perturbation directions \(\{{\bu}_{k,n}\}_{n=1}^N\) from the unit sphere \(\mathbb{S}_{d-1}\) to estimate the directional gradients \(G_{f,k,n}^{(2)}\) using Equation \eqref{eq:TZO-Gra}. To counteract the high variance typical of ZO estimation, these directional gradients are averaged over a mini-batch of \(N\) directions, producing a stable, multi-point gradient estimator: $\hat{\bg}_k = \frac{1}{N} \sum_{n=1}^N G_{f,k,n}^{(2)}$, which will replace the $\hat{\bg}_k$ in Equation \eqref{eq:AdaMM_base}. Furthermore, to ensure the updated control parameters remain within their physically safe operational domain \(\mathcal{X}\), we apply a projection operator \(\Pi_\mathcal{X}\), which enforces constraints to maintain system feasibility, as we describe in Section \ref{sec:P_F}.

\begin{algorithm}[ht]
\caption{Projected Multi-Point ZO-Adam (PMZO-Adam) for Coordinated IBR Control Optimization.}
\label{alg:PMZO-AdaMM_Unified}
\begin{algorithmic}[1]
\STATE \textbf{Input:} Disturbance scenario set $\mathcal{S}$, batch size $N$, max iteration number $K$,  tolerance $\tau$, Adam parameters $\beta_1, \beta_2 \in [0,1)$, $\epsilon \!>\! 0$, decay rates $\gamma_\eta, \gamma_r \in (0, 1)$, and minimum bounds $\eta_{\min}, r_{\min}$.
\STATE \textbf{Initialize:} IBR control parameters $\bx_1 \in \mathcal{X}$, moments $\bmm_0 = \textbf{0}$, $\bv_0 = \textbf{0}$, step size $\eta_1$, smoothing radius $r_1$.
\FOR{$k = 1, 2, \ldots, K$}
    \STATE \textbf{(1) Perform Parallel Simulations:} 
    \FOR{$n = 1, \ldots, N$ {in parallel}}
        \STATE Sample random direction $\bu_{k,n} \sim \text{Unif}(\mathbb{S}_{d-1})$. \label{step:sample}
        \STATE Execute simulations for each disturbance $\bd_s$ for $s\in\sS$ in parallel to obtain the objective function values $f(\bx_k\!+\! r_k \bu_{k,n})$ and $f(\bx_k\!-\! r_k \bu_{k,n})$.\label{step:simu}
        % \STATE Compute directional gradient $G_{f,k,n}^{(2)}$ using Eq. \eqref{eq:TZO-Gra}
    \ENDFOR
    \STATE \textbf{(2) Gradient Estimation:}
    \STATE Compute average multi-point ZO gradient estimate:
    \begin{align}\label{eq:multiZO}
    {\bg}_k \!=\! \frac{1}{N} \sum_{n=1}^N\! \Big[\frac{d}{2r_k}\big(f(\bx_k\!+\! r_k \bu_{k,n})-f(\bx_k\!-\! r_k \bu_{k,n})\big)\bu_{k,n}\Big]. 
    \end{align}
    
    \STATE \textbf{(3) Optimization Update:}
    \STATE Compute the moment estimates $\bmm_k, \bv_k, \hat{\bmm}_k, \hat{\bv}_k$ according to \eqref{eq:ZOAdaMM:m} and \eqref{eq:ZOAdaMM:up}, and update the iterate by: \label{step:update}
\begin{align}\label{eq:iterate}
    \bx_{k+1} = \text{Proj}_{\mathcal{X}}\Big( \bx_k - \eta_k \frac{\hat{\bmm}_k}{\sqrt{\hat{\bv}_k} + \epsilon}\Big).
\end{align}    
    % \STATE Update the next step, using $\bx_{k+1} = \Pi_{\bx_k\in\mathcal{X}}( \bx_k - \eta_k \frac{\hat{\bmm}_k}{\sqrt{\hat{\bv}_k} + \epsilon})$, $\Pi_{\bx\in\mathcal{X}}(\bz):=\arg \min_{\bx\in \sX}\|\bx-\bz\|^2_2$
    \STATE Update the stepsize and smoothing radius by: \label{step:szsr}
    \begin{align}\label{eq:szsr}
        \eta_{k+1} \!=\! \max(\gamma_\eta\cdot\eta_k,\eta_{\min}),r_{k+1} \!=\!\max(\gamma_r\cdot r_k,\,r_{\min}).
    \end{align} 
    \STATE \textbf{(4) Convergence Check:} 
    
   \STATE If $||\bx_{k+1}-\bx_k||\leq \tau$, terminate.
    % \IF{$|f_{k}-f_{k-1}|\le \tau$}
    %     \STATE \textbf{break}
    % \ENDIF
    % \chen{what is $f_k$? $f(\bx_k)$? then you need to run simulations again. You can change to $||\bx_{k+1}-\bx_k||\leq \tau$}
\ENDFOR
\STATE \textbf{Output:} Optimized IBR control parameters $\bx^* = \bx_{k+1}$.
\end{algorithmic}
\end{algorithm}

% The decision variables in $\bx$ directly map to the critical control parameters introduced earlier in Section \ref{sec:controller}. To ensure stable gradient estimation across disparate parameter scales, the algorithm normalizes the search space to $[0,1]^d$ and converts back to physical values during system evaluations. To guarantee convergence, both the learning rate $\eta_r$ and smoothing radius $\delta_r$ decay monotonically according to $\eta_{r+1} = \max(\eta_{\min}, \eta_r \cdot \gamma_\eta)$ and $\delta_{r+1} = \max(\delta_{\min}, \delta_r \cdot \gamma_\delta)$, enabling robust exploration initially while allowing precise fine-tuning near convergence.

\begin{remark}\normalfont
    In Algorithm \ref{alg:PMZO-AdaMM_Unified}, each iteration requires $2N\cdot|\sS|$ simulations to construct the multi-point gradient estimate $\bg_k$. Nevertheless, these simulations are independent and can be executed in parallel. With sufficient computational resources, the wall-clock time per iteration can approach that of a single simulation.
Moreover, the algorithm is intended for offline tuning using available power system simulators, consistent with industry practice in which IBR controller parameters are configured in advance and adjusted infrequently. Additionally, advanced simulation techniques, such as hybrid phasor-EMT simulations \cite{panigrahy2016real}, 
can further reduce computational costs and support application to large-scale power systems. 
\qed
\end{remark}

Convergence results have been established for closely related adaptive ZO algorithms, such as ZO-AdaMM \cite{chen2019zo}, under Lipschitz-continuous and bounded gradients and specified algorithmic settings. These results provide theoretical motivation for the proposed PMZO-Adam algorithm, 
although they do not directly establish its convergence. Rigorous convergence analysis for the present application is challenging due to the potentially nonsmooth nonconvex transient objective, the Adam updates, and the customized schemes on the stepsize and smoothing radius. Accordingly, this paper assesses the empirical convergence performance of Algorithm \ref{alg:PMZO-AdaMM_Unified}. Section~\ref{sec:simulation} demonstrates substantial objective reductions and stabilization across several tested cases. These results support the algorithm's practical effectiveness, while a rigorous convergence analysis is left for future work.

\section{Simulation Tests and Analysis}\label{sec:simulation}

This section evaluates the proposed PMZO-Adam algorithm using high-fidelity EMT simulations under two representative disturbances: a sudden load change and a line contingency. The algorithm’s effectiveness is then assessed across multiple contingency scenarios. Finally, the effects of the batch size $N$ and the Adam method are studied.

% through several dynamic simulations. We first detail the simulation setup, followed by a demonstration of the algorithm's performance under two distinct transient disturbances, including load change and line contingency.  Finally, we provide an algorithmic analysis examining the influence of batch sizes and the impact of the AdaMM.
% \chen{add a summary paragraph to introduce what will be presented}

\subsection{Simulation Setup}

\begin{figure}
    \centering
    \includegraphics[width=0.9\linewidth]{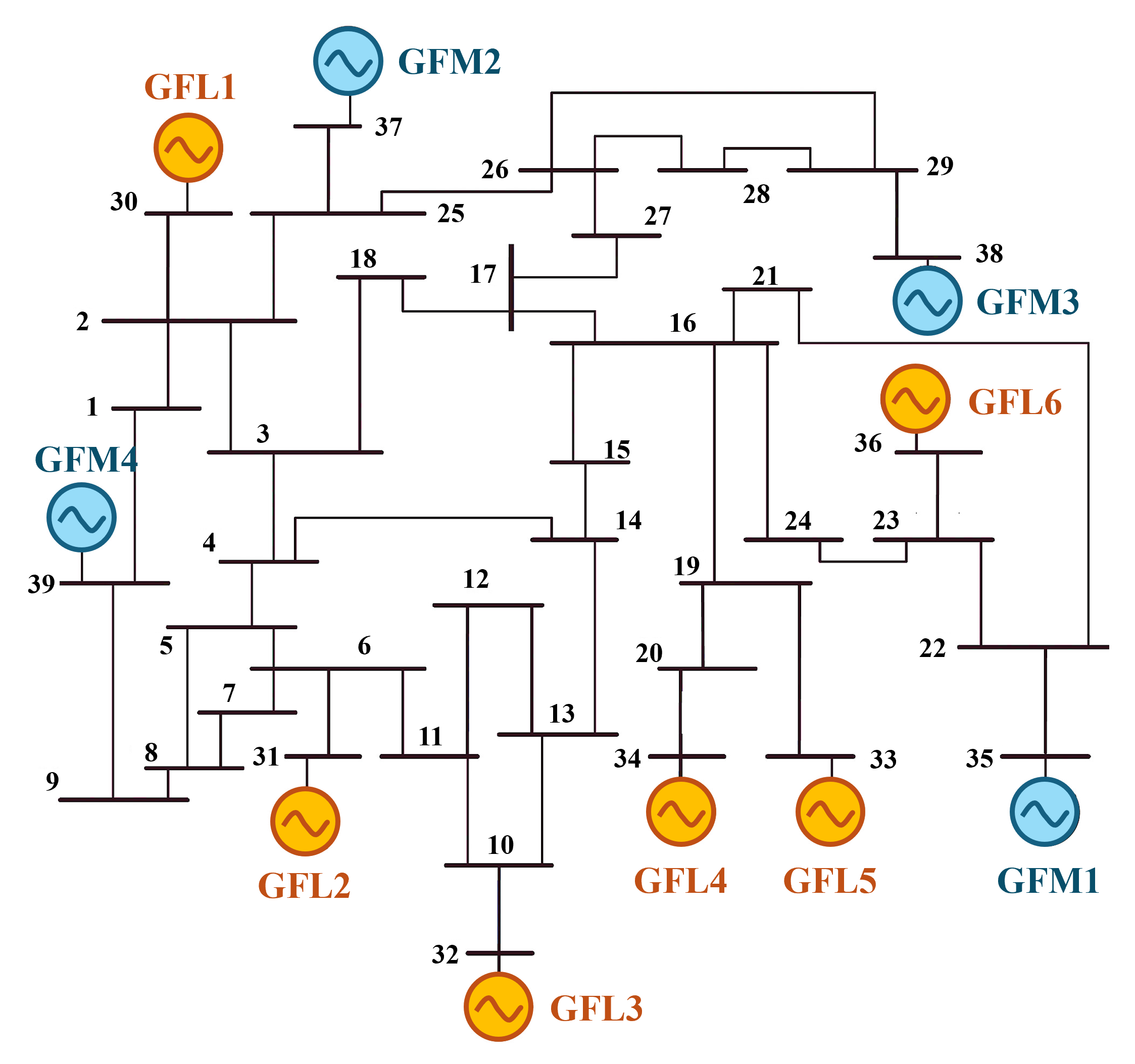}
    \caption{The modified IEEE 39-bus power system for simulation tests.}
    \label{fig:ieee39}
\end{figure}

A high-fidelity EMT dynamic model for a modified IEEE 39-bus system
is developed in MATLAB Simulink and used as the test system. As shown in Figure \ref{fig:ieee39}, the system includes six GFL and four GFM IBRs, each equipped with realistic control loops as described in Section \ref{sec:controller}. The detailed system parameters, initial control settings, the feasible set $\mathcal{X}$, and the hyperparameters of the PMZO-Adam algorithm are provided in  Appendix \ref{appendix}. The buses connected to IBRs are selected as the set $\sN_{\text{opt}}$ in \eqref{eq:objective}, whose transient frequency behaviors are considered for optimization. Since only primary frequency control is implemented, the system can stabilize the grid frequency after disturbances while not restoring the nominal frequency value. Below, we consider two representative disturbance cases, i.e., sudden load change (Case 1) and line contingency (Case 2), and optimize the IBR control parameters separately for each case to enhance the grid transient frequency performance as defined in \eqref{eq:objective}.  Then, Case 3 considers multi-scenario control parameter optimization according to \eqref{eq:base_obj}. 

The simulations are implemented in MATLAB R2025a on a computing platform equipped with an AMD Ryzen 5 8400F CPU, an NVIDIA GeForce RTX 5060 GPU, and 48 GB of RAM. The EMT Simulink model is executed with a time step of $50 \mu$s. Each simulation spans $5$ seconds and takes approximately $40$ seconds on average to complete.

%\chen{[\checkmark] I think it would be better to add the simulation efficiency. So please add the following information.}
% The simulations are implemented in MATLAB R2025a under a computing environment with an AMD Ryzen 5 8400F CPU, NVIDIA GeForce RTX 5060 GPU, and 48GB of RAM. The EMT power system model is executed with a time step of $50 \mu s$. Each simulation runs for $5$ seconds and takes approximately $40$ seconds on average to complete.  %\chen{[\checkmark] 48GB is RAM?\textit{ yes}.}

% The numerical case studies utilize a modified IEEE 39-bus system, as illustrated in Figure \ref{fig:ieee39}. The system includes 6 GFL and 4 GFM inverters. The detailed power parameters, initial control settings, the feasible parameter set $\mathcal{X}$, and the hyperparameters for the PMZO-AdaMM algorithm are provided in the Appendix. The transient model was constructed and simulated in Simulink with a time step of $50 \mu s$, and the algorithm was executed in MATLAB R2025a.
% \chen{what are these three cases? saying you optimize the IBR control parameters for each case}

\subsection{Case 1: Sudden Load Change}\label{subsec:Load Change}

\begin{figure}
    \centering
    \includegraphics[width=1\linewidth]{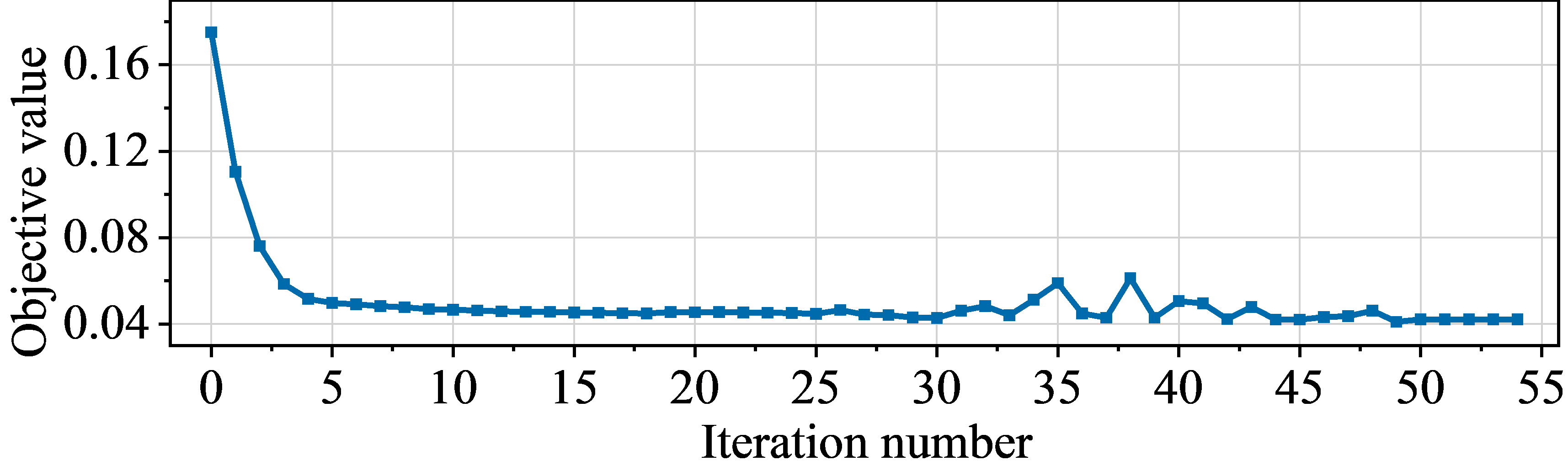}
    \caption{The convergence of objective function value in Case 1.} %\chen{To save space, you can squeeze the height of this figure a bit.}}
    \label{fig:lossall1}
\end{figure}

\begin{figure}
    \centering
    \includegraphics[width=1\linewidth]{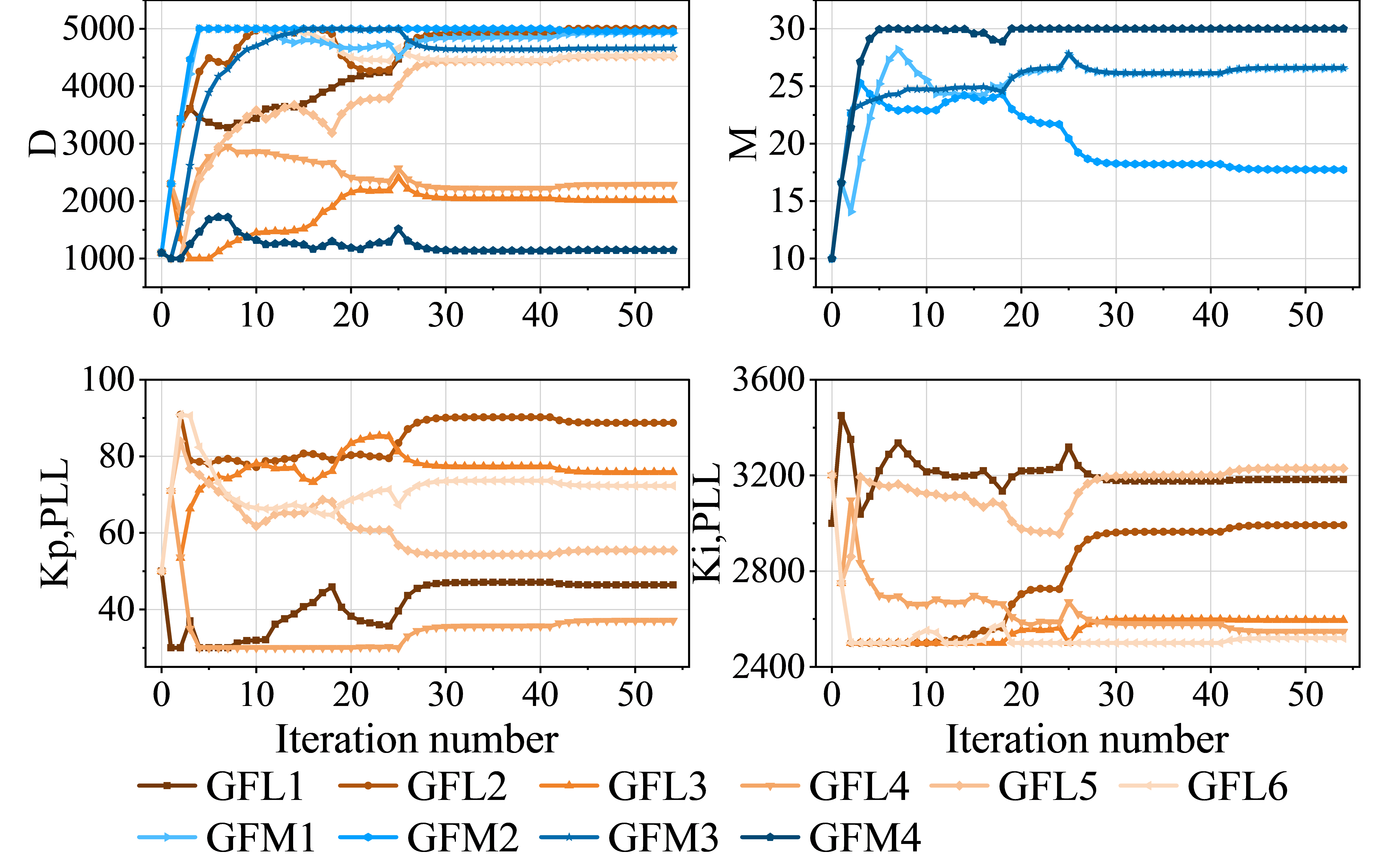}
    \caption{The convergence of key IBR control parameters in Case 1.} %\chen{for the legends, you can remove IBRx, just use GFL1, GFL2, xxx, GFM1, xxx, also enlarge "Iteration number" a bit }}
    \label{fig:para_trend1}
\end{figure}

\begin{figure}
    \centering
    \includegraphics[width=1\linewidth]{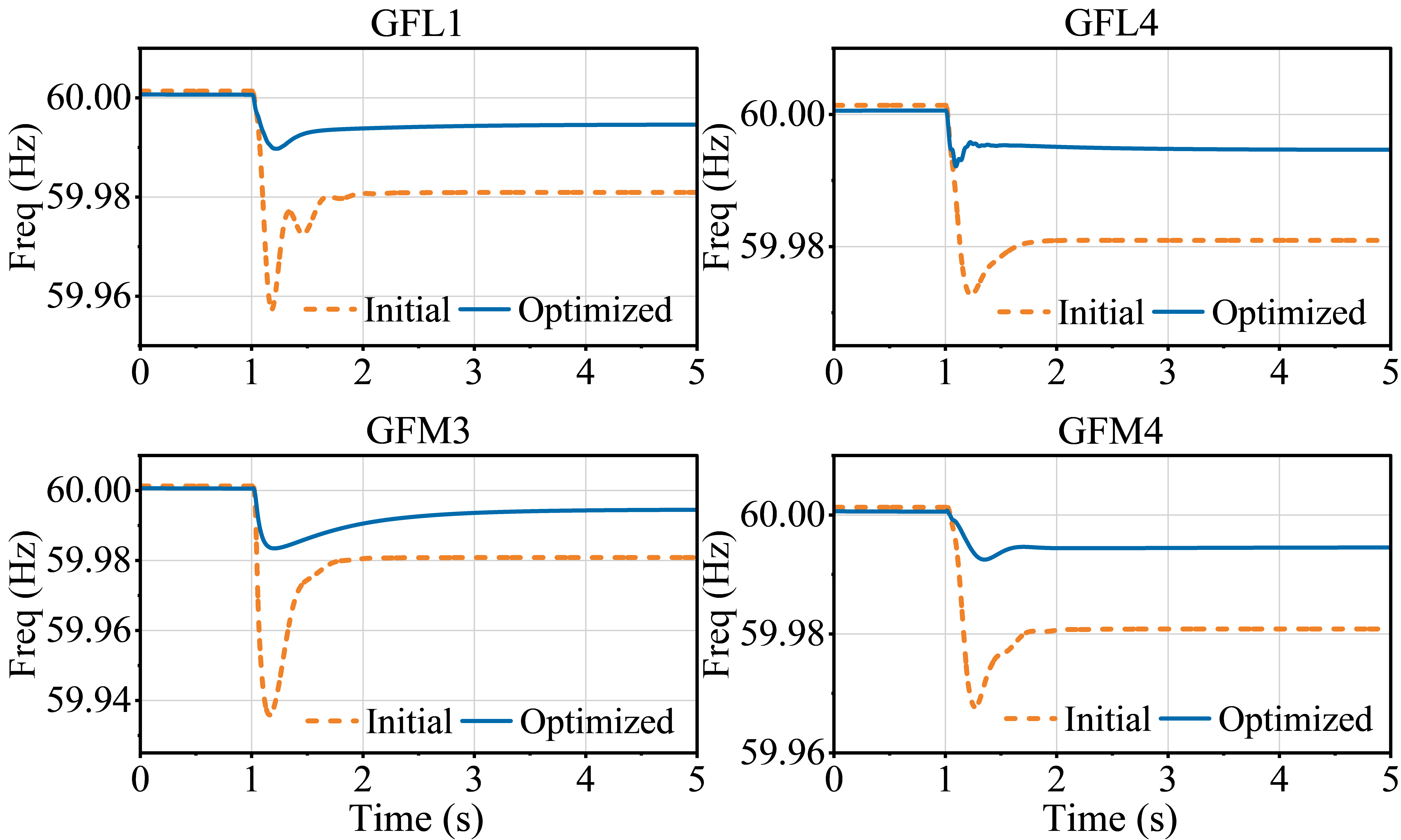}
    \caption{Frequency responses of four IBR buses under Case-1 disturbance, using the IBR control parameters before and after optimization. %\chen{no need the red dashed lines, change "Final" to "optimized", capitalize "Case"}
    }
    \label{fig:freq_respon1}
\end{figure}

In this subsection, we consider a large load increase disturbance that causes a significant drop in grid frequency. Specifically, an additional 1 GW load is suddenly applied at bus 26 at time $t \!=\! 1$s, while the total load in the system is 6 GW. The proposed algorithm is then employed to optimize the IBR control parameters. The optimization results are presented in Figures \ref{fig:lossall1}, \ref{fig:para_trend1}, and \ref{fig:freq_respon1}, as well as in Table \ref{tab:Opt_con_para}. 
%\chen{[\checkmark] the added load is 3.3MW? what is the total load in the system?} \chen{change the simulation time to 0-5 seconds, and t=1s for disturbance. \textit{I used the original value of load here, which is 1GW}}

%We focus on the frequency of all buses connected with IBRs. Optimization results are depicted in Figures \ref{fig:lossall1}, \ref{fig:para_trend1}, \ref{fig:freq_respon1}, and Tables \ref{tab:Opt_con_para}.

Figure \ref{fig:lossall1} illustrates the convergence of the objective function value, which decreases from 0.1751 to 0.0421, indicating a 76.0\% improvement in grid frequency transient performance after optimizing the IBR control parameters. Moreover, a rapid decrease in the objective value is observed within the first 10 iterations, demonstrating the fast convergence of the proposed algorithm. The small fluctuations in subsequent iterations arise from the random perturbations for gradient estimation in the algorithm. Figure \ref{fig:para_trend1} shows the corresponding convergence of the key IBR control parameters, i.e., the decision variables $\bx$.

The grid transient frequency responses under the Case-1 disturbance, before and after optimization, are compared and illustrated in Figure \ref{fig:freq_respon1}. We present the frequency dynamics of the buses connected to GFL1, GFL4, GFM3, and GFM4, as they are most affected by the disturbance. It is observed that both the frequency nadirs and fluctuations are significantly improved when using the optimized IBR control parameters generated by the proposed algorithm. For example, GFM3 is located close to the load disturbance at bus 26 and exhibits a substantial frequency response; after optimization, both its damping and inertia parameters are increased to provide stronger local support and mitigate overall grid frequency deviations.
These results demonstrate that the proposed algorithm effectively optimizes the IBR control parameters in response to load disturbances, leading to significant improvements in grid transient performance.

\subsection{Case 2: Line Contingency} \label{subsec:Line Contingency}

\begin{figure}
    \centering
    \includegraphics[width=1\linewidth]{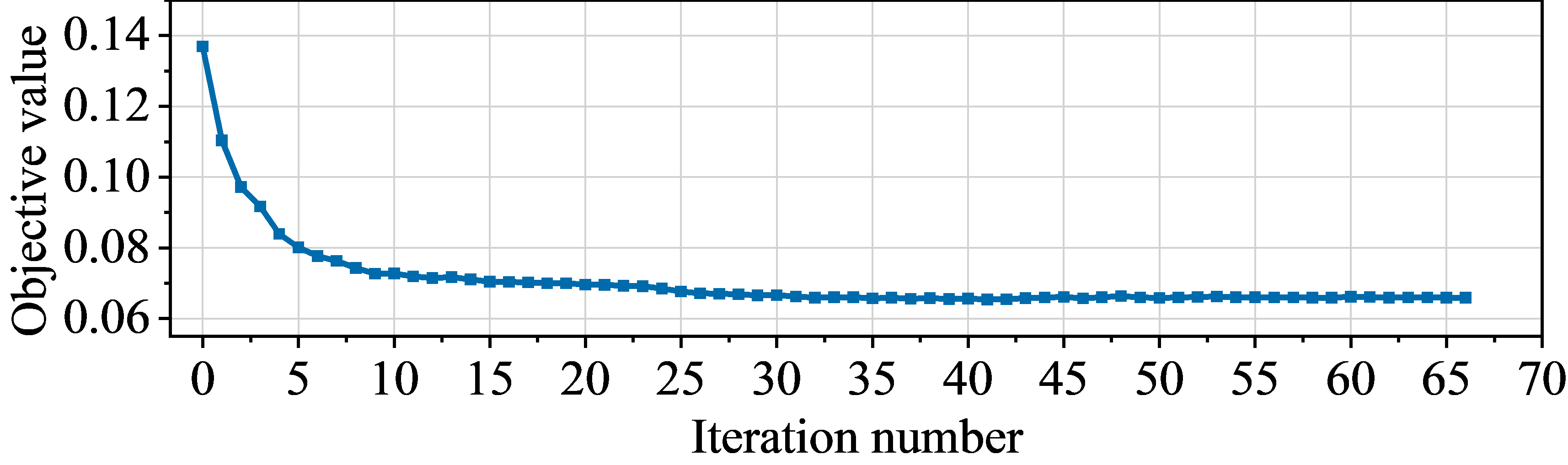}
    \caption{The convergence of objective function value in Case 2.}
    \label{fig:linecontin_loss}
\end{figure}

\begin{figure}
    \centering
    \includegraphics[width=1\linewidth]{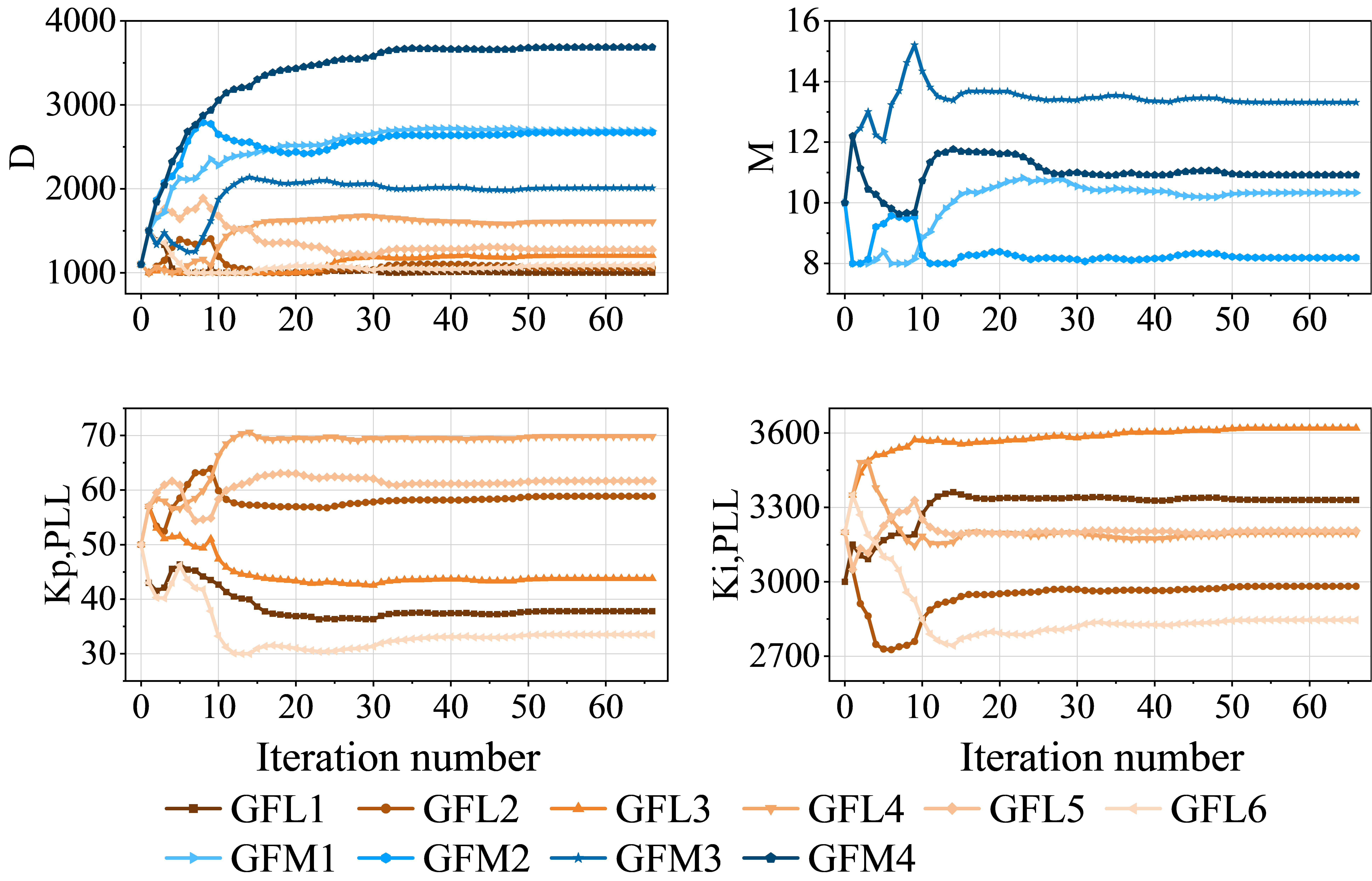}
    \caption{The convergence of key IBR control parameters in Case 2. %The convergence of important control parameters of Case 2.
    }
    \label{fig:para_con2}
\end{figure}

\begin{figure}
    \centering
    \includegraphics[width=1\linewidth]{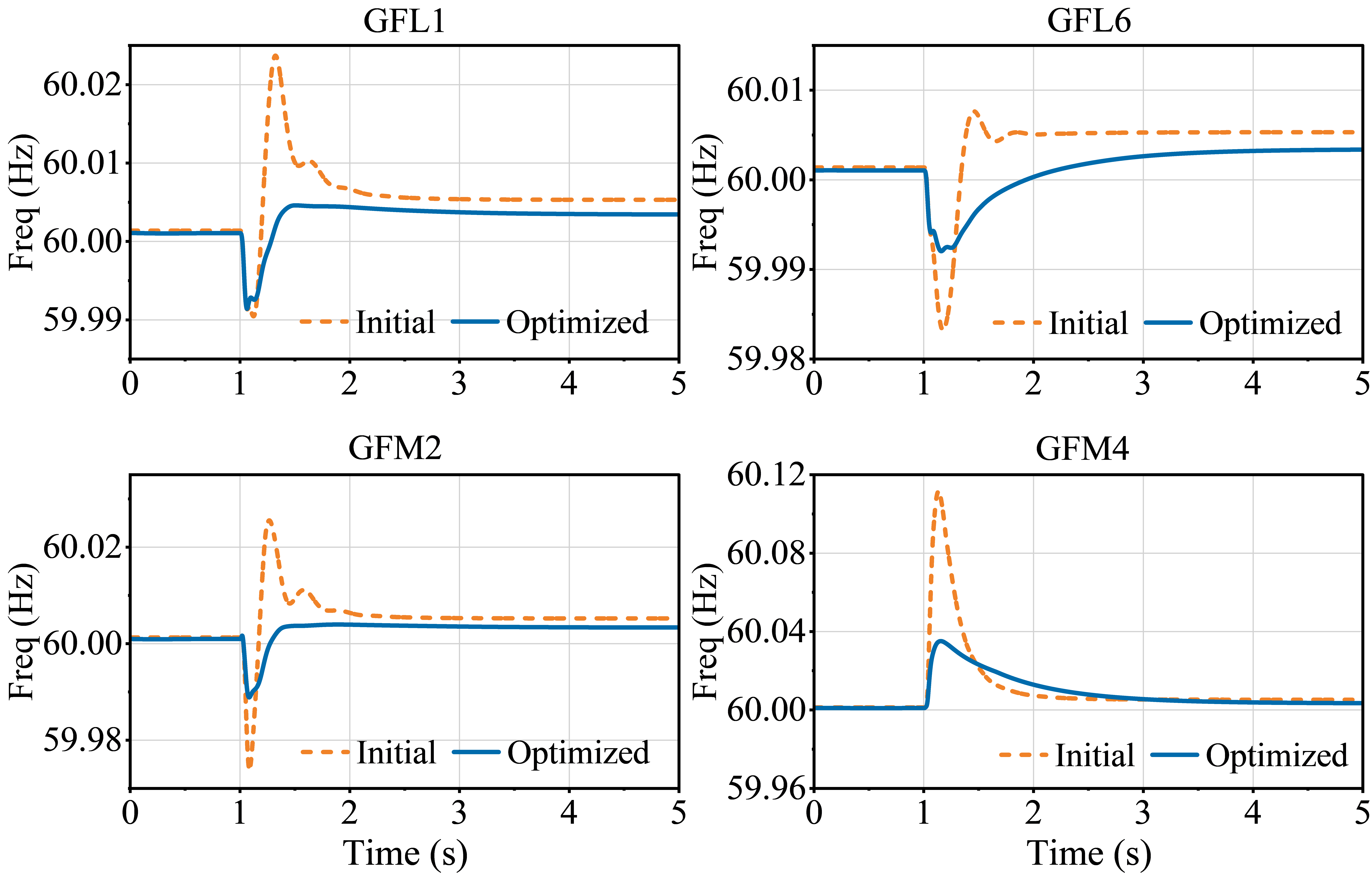}
    \caption{Frequency responses of four IBR buses under Case-2 disturbance, using the IBR control parameters before and after optimization.}
    \label{fig:freq_response2}
\end{figure}

In this subsection, we consider a severe line contingency disturbance in which the transmission line between bus 8 and bus 9 is suddenly opened at $t\!=\!1$s, which alters the network topology and causes large frequency deviations. The results are presented  in Figures \ref{fig:linecontin_loss}, \ref{fig:para_con2}, \ref{fig:freq_response2}, and Table \ref{tab:Opt_con_para2}.

% due to the disconnection of a transmission line, which affects system topology and causes rapid frequency deviations. Specifically, the line between bus 18 and bus 19 opens at $t=6s$. Optimization results are shown in Figures \ref{fig:linecontin_loss}, \ref{fig:para_con2}, \ref{fig:freq_response2}, and Tables \ref{tab:Opt_con_para2}. 

Figure \ref{fig:linecontin_loss} illustrates the objective function value converging from 0.1369 to 0.066 (a 51.8\% improvement), with a rapid decrease observed within the first 10 iterations. Figure \ref{fig:para_con2} shows the convergence of the key IBR control parameters during the optimization process. Figure \ref{fig:freq_response2} compares the frequency responses of the four most affected IBR buses, showing that maximum frequency deviations and oscillations are significantly mitigated after optimization. For instance, the damping parameter $D$ of GFM4 (closest to the fault) is substantially increased to improve transient behavior, while $K_{p,\text{PLL}}$ of GFL1 is reduced to prevent excessive overshoot. This highlights the necessity of coordinated parameter tuning across grid IBRs for effective disturbance mitigation.

% Figure \ref{fig:linecontin_loss} indicates strong convergence, with the objective value reduced by 51.8\% from 0.1369 to 0.0660. Figure \ref{fig:para_con2} shows parameter convergence during optimization, while Figure \ref{fig:freq_response2} highlights significant performance enhancements. Unlike uniform load increases, a line contingency alters the grid's admittance matrix, affecting power distribution. The coordinated optimization effectively redistributed damping responsibilities. GFM4, closest to the fault, experienced considerable improvement post-optimization, with its damping parameter $D$ increasing more than any other IBR, playing a key role in minimizing the objective value. Similarly, GFL1's performance improved with increased $D$ and reduced $K_{p,\text{PLL}}$, consistent with Case 1, highlighting the importance of coordinated parameter adjustments across IBR types for effective disturbance mitigation.

The case studies above demonstrate the robustness and effectiveness of the proposed method in handling different types of disturbances. The method can also be used to identify optimal IBR control parameters that balance transient dynamic performance across multiple disturbance scenarios.

% These case studies demonstrate the robust performance of the proposed method across various disturbance scenarios, ensuring reliable convergence. The optimized settings significantly enhance the capabilities of individual IBRs and the overall system to withstand disturbances while maintaining frequency stability.  


\subsection{Case 3: Multiple Contingencies} 
\begin{figure}
    \centering
    \includegraphics[width=1\linewidth]{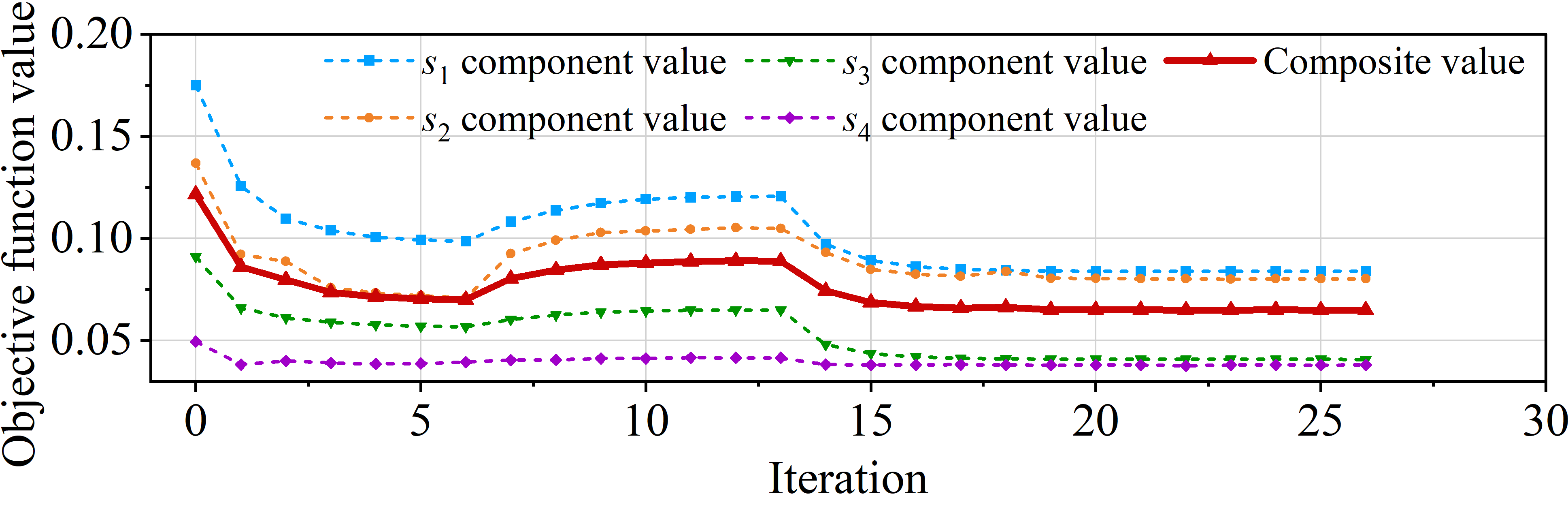}
    \caption{The convergence of the individual scenario objectives and their weighted sum
    in Case 3, considering the load change scenarios $s_1, s_3$ and the line contingency scenarios $s_2, s_4$.
    %. The composite value includes the component value of the load change scenario $s_1$ and the component value of the line contingency scenario $s_2$.}
    %The convergence of important control parameters of Case 2.
    }
    \label{fig:object_multi}
\end{figure}

\begin{figure}
    \centering
    \includegraphics[width=1\linewidth]{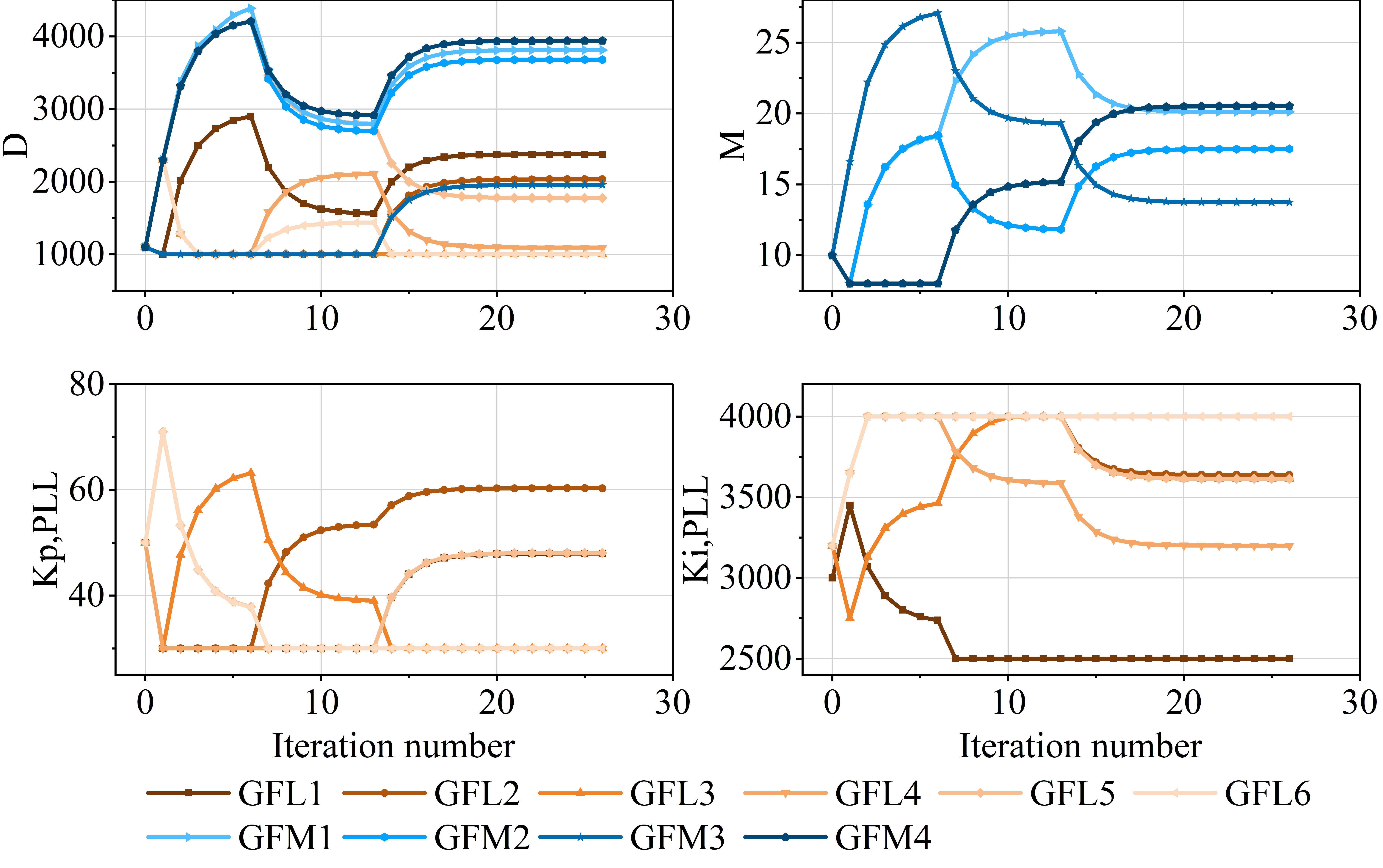}
    \caption{The convergence of key IBR control parameters in Case 3.
    %The convergence of important control parameters of Case 2.
    }
    \label{fig:para_con3}
\end{figure}

\begin{figure}
    \centering
    \includegraphics[width=1\linewidth]{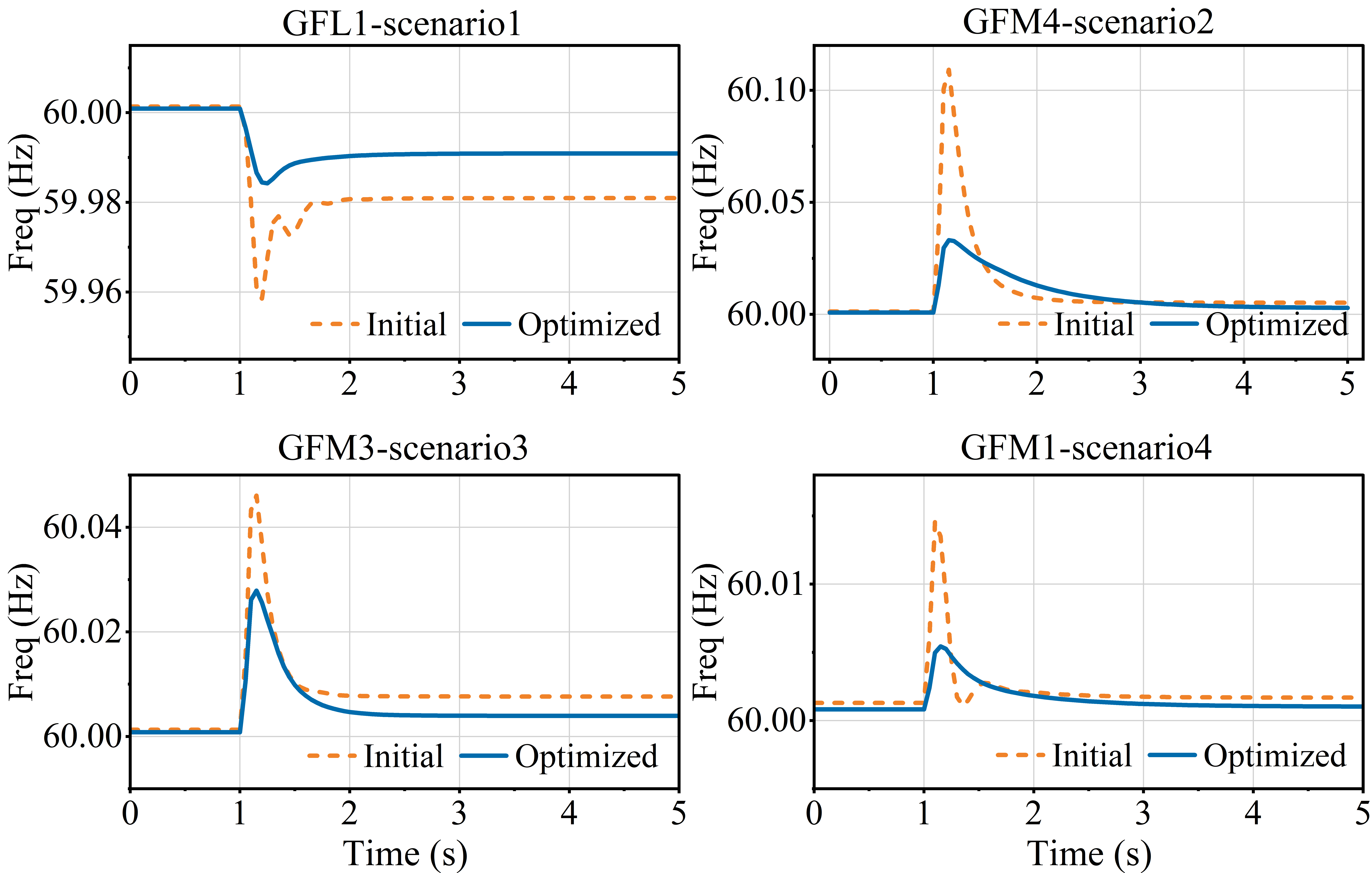}
    \caption{Frequency responses of IBR buses of interest under Case 3 multiple disturbances, using the IBR control parameters before and after optimization. %\chen{no need the red dashed lines, change "Final" to "optimized", capitalize "Case"}
    }
    \label{fig:freq_respon3}
\end{figure}

    To evaluate the framework across multiple disturbance scenarios, this subsection considers a scenario set $\sS=[s_1,s_2,s_3,s_4]$, comprising $s_1$ (sudden load increase, Case 1), $s_2$ (line contingency, Case 2), $s_3$ (sudden load decrease, bus 29), and $s_4$ (another line contingency, bus 14-15). In each optimization step, independent EMT simulations are performed for the four scenarios, with each disturbance applied separately.
    Following \eqref{eq:base_obj}, 
    the scenario objectives are combined using weights $\alpha_s=[0.3,0.3,0.2,0.2]$, placing greater emphasis on $s_1$ and $s_2$. The simulation results are presented in Figures \ref{fig:object_multi}, \ref{fig:para_con3}, \ref{fig:freq_respon3}, and Table \ref{tab:Opt_con_para3}.

    % for each candidate parameter setting, EMT simulations are performed independently for each scenario $s$, which means these contingencies do not happen simultaneously. The objective values are aggregated via a weighted sum with contingency weight coefficients $\bm{\alpha}_s = [0.6, 0.4]^\top$, placing higher priority on the dynamic response under load variations. The resulting composite objective is then used to generate gradient estimates for parameter updates. The results are presented in Figures \ref{fig:object_multi}, \ref{fig:para_con3}, \ref{fig:freq_respon3}, and Table \ref{tab:Opt_con_para3}.

    Figure \ref{fig:object_multi} shows the evolution of the individual scenario objectives and their weighted sum. The composite objective value decreases from 0.1218 to 0.0651, a reduction of 46.6\%. Notably, while the optimization is primarily driven by the heavily weighted and more severe scenarios ($s_1$ and $s_2$), the performance in the less severe scenarios ($s_3$ and $s_4$) does not degrade, but rather remains stable or slightly improves. This indicates that the weighted composite objective effectively prevents the controllers from overfitting to a single disturbance type. Figure \ref{fig:para_con3} illustrates the convergence of the key IBR control parameters. Unlike single-scenario optimization, where parameters might be aggressively tuned to counteract one specific fault, the multi-scenario framework forces the parameters to stabilize at a robust balance. For instance, the damping parameters $D$ of the most affected IBRs are increased to provide vital frequency support, but they are implicitly constrained by the algorithm from reaching extreme values that might trigger instability or excessive overshoot in other contingency scenarios. Figure \ref{fig:freq_respon3} further compares the frequency responses of several IBR buses before and after optimization. The results confirm that a single, universally coordinated parameter set can effectively mitigate maximum frequency deviations and dampen oscillations across diverse network topologies and active power imbalance conditions. Ultimately, this demonstrates that the proposed model-free framework is capable of deriving a highly robust controller setting, guaranteeing grid-level transient stability under a wide spectrum of unpredictable real-world disturbances.

\subsection{Impact of Batch Size $N$}

% \chen{[\checkmark] what is the batch size in the cases above?\textit{ Actually, they both used 2. i added the description in the paragraph below.}}

In the PMZO-Adam algorithm (Algorithm \ref{alg:PMZO-AdaMM_Unified}), a multi-point zeroth-order gradient estimator with $N$ batches is employed to reduce variance and enhance convergence. In the previous tests above, the batch size $N$ was set to 2. In this subsection, we examine the impact of the batch size $N$ on the algorithm using Case 1 as the test case. 

\begin{figure}
    \centering
    \includegraphics[width=1\linewidth]{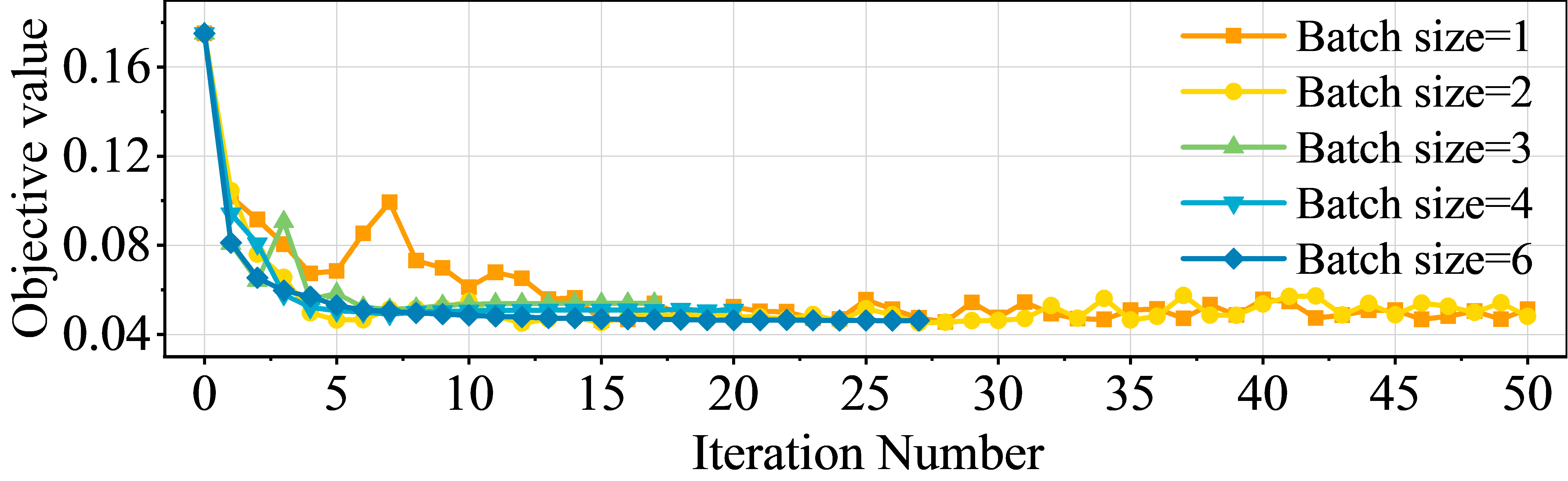}
    \caption{The convergence of objective value under different batch sizes.}%\chen{[\checkmark] This figure is very hard to distinguish. You can use different colors for different batch sizes, also enlarge the data markers, now are too small to see} }
    \label{fig:batch_test}
\end{figure}

Figure \ref{fig:batch_test} illustrates the convergence of the objective function value for different batch sizes $N$. Small batch sizes ($N = 1, 2$) exhibit significant fluctuations due to the high variance of the gradient estimates, with $N = 1$ showing a prolonged exploration phase before stabilization. As $N$ increases, the convergence curves become smoother and faster. In particular, a batch size of $N = 6$ yields the most favorable convergence trajectory with the steepest descent and the lowest final objective value. This observation is consistent with the intuition that a larger batch size $N$ generally reduces the gradient estimation variance and leads to more stable convergence. Although a larger $N$ requires more simulation evaluations, these evaluations can be executed fully in parallel within the proposed framework. Nevertheless, in highly nonconvex optimization landscapes, the large variance associated with small batch sizes can facilitate broader exploration and occasionally help the algorithm escape local minima \cite{ren2023escaping}. Choosing the optimal batch size is key to balancing stability and exploration, which remains for future work.

% In Algorithm \ref{alg:PMZO-AdaMM_Unified}, the batch size $N$ determines the number of random directions sampled for gradient estimation in each iteration. This hyperparameter plays a crucial role in influencing convergence speed and gradient estimation accuracy. Generally, a larger $N$ reduces variance in gradient estimates, allowing for more stable convergence, whereas a smaller $N$ produces noisier gradients. For this test, with the PMZO-AdaMM algorithm, the maximum iteration count is set to 50, using Case 1 as the disturbance scenario.

% Figure \ref{fig:batch_test} illustrates the impact of varying batch sizes $N$ on convergence behavior. Smaller batch sizes ($N = 1, 2$) exhibit significant fluctuations due to high variance in gradient estimation, with $N = 1$ undergoing prolonged exploration phases before stabilizing. As $N$ increases, convergence curves smooth out and proceed faster. A batch size of $N=6$ yields the most favorable convergence trajectory, demonstrating the sharpest descent and the lowest final objective value. Nonetheless, while some larger batch sizes ($N=3, 4$) facilitate quicker convergence, in highly non-convex landscapes, the inherent gradient noise of smaller batches can occasionally act as a regularizer, helping the algorithm escape local minima \cite{ren2023escaping}. Therefore, choosing an optimal batch size is essential to balancing convergence stability, speed, and exploration capacity.

\begin{figure}
    \centering
    \includegraphics[width=1\linewidth]{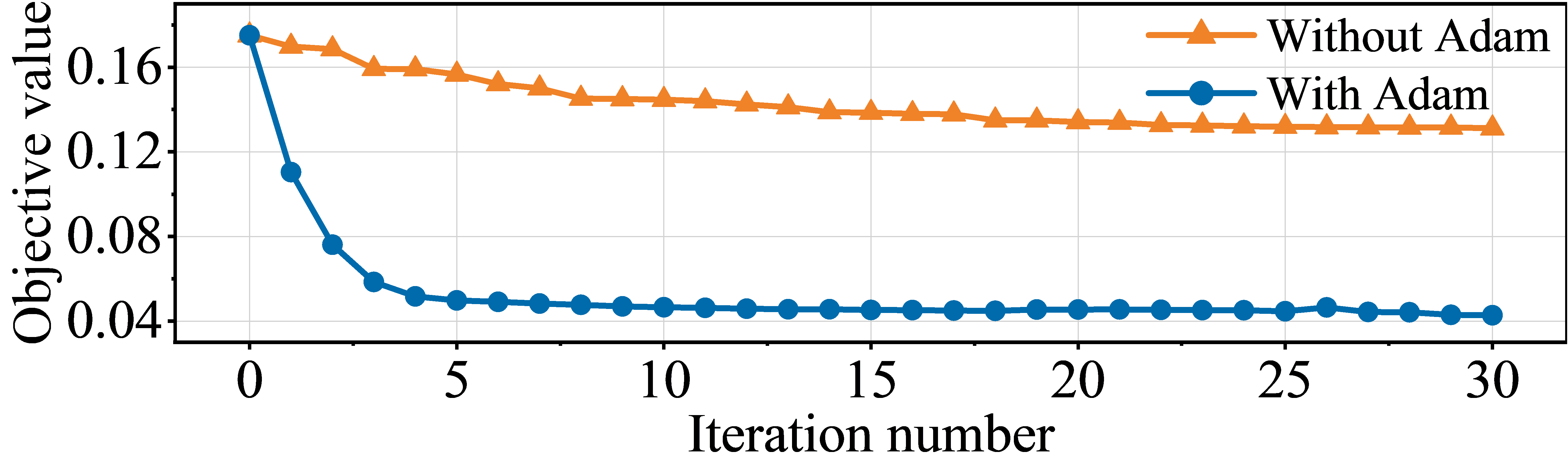}
    \caption{The convergence of objective value with and without Adam. %\chen{change the legends in the figure to be Without Adam and With Adam}
    }
    \label{fig:With and without AdaMM}
\end{figure}

\subsection{Effect of the Adam Method}

As discussed in Section \ref{sec:zo}, the Adam method improves convergence through adaptive moment mechanisms. This subsection examines the effect of incorporating Adam into the proposed algorithm. The tests are conducted under Case 1 disturbance with a batch size of $N = 2$ and an initial stepsize of $\eta_1 = 0.1$. Figure \ref{fig:With and without AdaMM} presents the ablation study on the Adam method. The proposed PMZO-Adam algorithm exhibits significantly faster convergence and achieves a lower final objective value than the variant without Adam. This result demonstrates the effectiveness of incorporating Adam to accelerate convergence, which adaptively updates the stepsize and descent direction based on gradient history.

% As discussed in Section \ref{sec:zo}, the AdaMM optimizer enhances parameter updates through adaptive momentum mechanisms. This section examines the effect of integrating AdaMM into the proposed algorithm. Tests are conducted under Case 1 with a batch size $N=2$ and step size $\eta_1=0.1$. Figure \ref{fig:With and without AdaMM} demonstrates the ablation study regarding the AdaMM optimizer. The proposed PMZO-AdaMM algorithm exhibits vastly accelerated convergence and achieves superior final objective values compared to the version without AdaMM. This improvement is attributed to AdaMM's adaptive momentum mechanisms, which efficiently maintain per-parameter learning rates that automatically adjust based on the gradient history.

\section{conclusion}\label{sec:conclusion}

This paper introduces a simulation-based, model-free 
framework for the coordinated optimization of IBR control parameters to enhance grid-level transient dynamic performance. The proposed PMZO-Adam algorithm solves the optimization problem in a model-free manner, eliminating the need for explicit mathematical models of complex nonlinear power system dynamics. Extensive EMT simulations on a modified IEEE 39-bus system demonstrate that the proposed approach significantly mitigates maximum frequency deviations and frequency fluctuations under large disturbances. In addition, incorporating multi-point ZO and the Adam method effectively reduces gradient estimation variance and accelerates convergence. Future work will focus on providing a theoretical performance analysis of the proposed algorithm and extending the framework to online settings and applications with more realistic constraints.

% method utilizing the PMZO-Adam algorithm to advance the optimization of IBR control parameters for improved grid-level transient frequency stability. Validated through case studies on a modified IEEE 39-bus system, the proposed approach effectively mitigates frequency dips during significant disturbances such as sudden load changes and line contingencies. The inclusion of Adam enhances convergence speed and resilience. While initial results are promising, this study primarily relies on simulation-based assessments. Future work aims to expand the theoretical grounding by integrating small-signal analysis, providing deeper insights into system dynamics and constraint interactions.

% \chen{revise the conclusion accordingly}

\appendices

\section{Simulation Settings}\label{appendix}

%\chen{Please edit the tables to save space and fit the 10 page limit. For example, not all of them needs a long table, e.g. Table III can be a short table in a single column. Some unimportant tables may be removed}

This appendix presents the simulation settings for the modified IEEE 39-bus system shown in Figure \ref{fig:ieee39}. Table \ref{tab:combined_params} summarizes the initial power flow data and baseline control parameters for all IBRs. Table \ref{tab:bounds_params} defines the feasible parameter set $\mathcal{X}$ by specifying the physical lower and upper bounds of these control parameters. Table \ref{tab:hyper_params} provides the hyperparameter settings used in the PMZO-Adam optimization algorithm. Tables \ref{tab:Opt_con_para}, \ref{tab:Opt_con_para2} and \ref{tab:Opt_con_para3} present the optimized parameter settings for Case 1, Case 2, and Case 3, respectively. 

% This appendix details the configurations for the modified IEEE 39-bus system simulations (see Figure \ref{fig:ieee39}). Table \ref{tab:combined_params} lists the initial power flow and baseline control parameters for all IBRs. Table \ref{tab:bounds_params} defines the feasible parameter set $\mathcal{X}$ by specifying the physical upper and lower bounds for these controls. Finally, Table \ref{tab:hyper_params} provides the hyperparameter settings used for the PMZO-Adam optimization algorithm. 
% Table \ref{tab:Opt_con_para} and \ref{tab:Opt_con_para2} are the optimal parameter settings of Scenario 1 and Scenario 2, respectively. 

\begin{table*}[t]
\centering
\caption{Inverter-Based Resources Parameters}
\label{tab:combined_params}
\begin{tabular*}{\textwidth}{@{\extracolsep{\fill}} c| ccc | cccccccc  @{}}
\toprule
\multicolumn{1}{c|}{}  & \multicolumn{3}{c|}{\textbf{IBR Information}} &\multicolumn{8}{c}{\textbf{Initial Control Parameters Settings}} \\
\midrule
\textbf{IBR} & Bus &  $P_{\text{init}}\ (\text{MW})$ & $Q_{\text{init}}\ (\text{MVar})$ & $D$ & $M$ & $K_{i,\text{PLL}}$ & $K_{p,\text{PLL}}$ & $K_{i,v}$ & $K_{p,v}$ & $K_{i,i}$ & $K_{p,i}$ \\
\midrule
GFL1  & 30 & 2 & 0.1  & 1100   & --  & 3000 & 50 & --   & -- & 200 & 50 \\
GFL2  & 31 & 2 & 0.1  & 1100   & -- & 3200 & 50 & --   & -- & 200 & 50 \\
GFL3  & 32 & 2 & 0.1 & 1100   & -- & 3200 & 50 & --   & -- & 100 & 20  \\
GFL4  & 34 & 2 & 0.1 & 1100   & -- & 3200 & 50 & --   & -- & 200 & 50  \\
GFL5  & 33 & 2 & 0.1 & 1100   & -- & 3200 & 50 & --   & -- & 200 & 70 \\
GFL6  & 36 & 2 & 0.1 & 1100   & -- & 3200 & 50 & --   & -- & 200 & 50  \\
GFM1  & 35 & 3 & 0   & 1100 & 10 & --   & --   & 40 & 10  & 12.5 & 2.5 \\
GFM2  & 37 & 3 & 0   & 1100 & 10 & --   & --   & 40 & 10  & 12.5 & 2.5 \\
GFM3  & 38 & 4 & 0  & 1100 & 10 & --   & --   & 40 & 10  & 12.5 & 2.5  \\
GFM4 & 39 & 5 & 0   & 1100 & 10 & --   & --   & 40 & 10  & 12.5 & 2.5 \\
\bottomrule
\end{tabular*}
\end{table*}

\begin{table*}[t]
\centering
\caption{Feasible intervals of Key IBR Control Parameters}
\label{tab:bounds_params}
\begin{tabular*}{\textwidth}{@{\extracolsep{\fill}} c| cccccccc  @{}}
\toprule
\textbf{IBR} & $D$ & $M$ & $K_{i,\text{PLL}}$ & $K_{p,\text{PLL}}$ & $K_{i,v}$ & $K_{p,v}$ & $K_{i,i}$ & $K_{p,i}$ \\
\midrule
GFL1,2,3,4,5,6  & [1000, 5000]   & --  & [2500,4000] & [30,100] & --   & -- & [20, 300] & [20,70] \\
GFM1,2,3,4  & [1000, 5000]   & [8, 30] & --   & --   & [20, 60] & [5, 40]  & [8, 20] & [1, 5] \\
\bottomrule
\end{tabular*}
\end{table*}

\begin{table}[t]
\centering
\caption{Hyperparameters of the PMZO-Adam Algorithm} %\chen{modify the simulation time settings in this table}}
\label{tab:hyper_params}
\begin{tabular}{ cc | cc | cc }
\toprule
\textbf{Parameter} & \textbf{Value} & \textbf{Parameter} & \textbf{Value} & \textbf{Parameter} & \textbf{Value} \\
\midrule
$\eta_1$ & 0.1 & $K$ & 70 & $\tau$ & $10^{-4}$ \\
$r_1$ & 0.1 & $N$ & 2 & $t_{\mathrm{d}}\ (s)$ & 1 \\
$\gamma_{\eta}$ & 0.9 & $\beta_1$ & 0.5 & $T\ (s)$ & 5 \\
$\gamma_{r}$ & 0.95 & $\beta_2$ & 0.99 & $t_\mathrm{o}\ (s)$ & 4.5 \\
$\eta_{\min}$ & 0.001 & $\epsilon$ & $10^{-8}$ & $\lambda$ & 0.5 \\
$r_{\min}$ & 0.001 &  & & & \\
\bottomrule
\end{tabular}
\end{table}

\begin{table}[t]
\centering
\caption{Optimal IBR Control Parameters under Case 1 Disturbance.}
\label{tab:Opt_con_para}
\setlength{\extrarowheight}{-2pt}
\begin{tabular}{@{\extracolsep{\fill}} c| cccccccc  @{}}
\toprule
\textbf{IBR} & $D$ & $M$ & $K_{i,\text{PLL}}$ & $K_{p,\text{PLL}}$ & $K_{i,v}$ & $K_{p,v}$ & $K_{i,i}$ & $K_{p,i}$ \\
\midrule
GFL1  & 4980.2
   & --  & 3182.7 & 46.4 & --   & -- & 43.4 & 24.5 \\
GFL2  & 5000.0   & -- & 2992.1 & 88.7 & --   & -- & 208.4 & 61.6 \\
GFL3  & 2015.5   & -- & 2595.9 & 75.9 & --   & -- & 109.4 & 42.8 \\
GFL4  & 2283.8   & -- & 2547.1 & 37.1 & --   & -- & 288.0 & 49.0 \\
GFL5  & 4519.3   & -- & 3229.4 & 55.4 & --   & -- & 203.6 & 37.1 \\
GFL6  & 4535.3   & -- & 2520.4 & 72.3 & --   & -- &  38.6 & 55.7 \\
GFM1  & 4925.6   & 26.5 & --   & --   & 27.7 & 29.9 & 9.5 & 2.2 \\
GFM2  & 4954.6   & 17.7 & --   & --   & 31.9 & 5.6  & 19.4 & 3.2 \\
GFM3  & 4658.0   & 26.6 & --   & --   & 55.2 & 5.1  & 18.0 & 4.7  \\
GFM4 & 1146.9   & 30.0 & --   & --   & 21.3 & 40.0  & 8.0 & 2.0 \\
\bottomrule
\end{tabular}
\end{table}

\begin{table}[t]
\centering
\caption{Optimal IBR Control Parameters under Case 2 Disturbance.}
\setlength{\extrarowheight}{-2pt}
\label{tab:Opt_con_para2}
\begin{tabular}{@{\extracolsep{\fill}} c| cccccccc  @{}}
\toprule
\textbf{IBR} & $D$ & $M$ & $K_{i,\text{PLL}}$ & $K_{p,\text{PLL}}$ & $K_{i,v}$ & $K_{p,v}$ & $K_{i,i}$ & $K_{p,i}$ \\
\midrule
GFL1  & 1000.0
   & --  & 3330.1 & 37.8 & --   & -- & 300 & 37.3 \\
GFL2  & 1057.1   & -- & 2981.9 & 58.9 & --   & -- & 291.1 & 62.1 \\
GFL3  & 1206.7   & -- & 3619.9 & 43.8 & --   & -- & 110.7 & 20.3  \\
GFL4  & 1605.3   & -- & 3197.0 & 69.8 & --   & -- & 250.5 & 60.2  \\
GFL5  & 1273.9  & -- & 3205.8 & 61.7 & --   & -- & 296.9 & 65.8 \\
GFL6  & 1086.9   & -- & 2846.0 & 33.5 & --   & -- & 204.6 & 40.0  \\
GFM1  & 2692.2   & 10.3 & --   & --   & 44.3 & 5.0  & 10.4 & 2.7 \\
GFM2  & 2671.6   & 8.2 & --   & --   & 49.7 & 5.0 & 15.8 & 1.2 \\
GFM3  & 2009.1   & 13.3 & --   & --   & 25.8 & 18.3  & 8.1 & 3.1  \\
GFM4 & 3686.3   & 10.9 & --   & --   & 47.7 & 19.0  & 9.8 & 1.9 \\
\bottomrule
\end{tabular}
\end{table}

\begin{table}[t]
\centering
\caption{Optimal IBR Control Parameters under Multiple Scenarios of Disturbance in Case 3.}
\setlength{\extrarowheight}{-2pt}
\label{tab:Opt_con_para3}
\begin{tabular}{@{\extracolsep{\fill}} c| cccccccc  @{}}
\toprule
\textbf{IBR} & $D$ & $M$ & $K_{i,\text{PLL}}$ & $K_{p,\text{PLL}}$ & $K_{i,v}$ & $K_{p,v}$ & $K_{i,i}$ & $K_{p,i}$ \\
\midrule
GFL1  & 2376.9 & --   & 2500.0 & 47.9 & --   & --   & 204.1 & 20.0 \\
GFL2  & 2032.4 & --   & 3638.4 & 60.3 & --   & --   & 20.0  & 32.9 \\
GFL3  & 1000.0 & --   & 3613.8 & 30.0 & --   & --   & 251.7 & 64.3 \\
GFL4  & 1092.2 & --   & 3199.8 & 30.0 & --   & --   & 263.1 & 54.5 \\
GFL5  & 1775.9 & --   & 3613.8 & 48.0 & --   & --   & 227.7 & 62.3 \\
GFL6  & 1000.0 & --   & 4000.0 & 30.0 & --   & --   & 74.2  & 57.1 \\
GFM1  & 3813.1 & 20.1 & --     & --   & 36.9 & 19.9 & 8.0   & 1.6  \\
GFM2  & 3680.6 & 17.5 & --     & --   & 30.8 & 31.0 & 12.9  & 2.4  \\
GFM3  & 1957.2 & 13.7 & --     & --   & 29.9 & 11.8 & 20.0  & 3.3  \\
GFM4  & 3940.8 & 20.5 & --     & --   & 20.0 & 13.8 & 12.9  & 3.6  \\
\bottomrule
\end{tabular}
\end{table}

\FloatBarrier % Flushes all pending tables before starting the bibliography

\bibliographystyle{IEEEtran}
\bibliography{IEEEabrv,mybibfile}

\end{document}